\documentclass[11pt,letterpaper]{article}
\pdfoutput=1
\usepackage{jheppub}

\usepackage{feynmf}
\usepackage{booktabs}
\usepackage[english]{babel}
\usepackage{amsmath,amssymb,amsbsy,amstext, amsthm, simplewick}
\usepackage{hyperref}
\usepackage{graphicx}
\usepackage{amsfonts}
\usepackage{amssymb}
\usepackage[small]{caption}
\usepackage{upgreek}
\usepackage[titletoc]{appendix}

\usepackage{pstricks}
\usepackage{color}

\def\beq{\begin{equation}}
\def\eeq{\end{equation}}

\def\ifmath#1{\relax\ifmmode #1\else $#1$\fi}
\def\lsim{{~\raise.15em\hbox{$<$}\kern-.85em\lower.35em\hbox{$\sim$}~}}
\def\gsim{{~\raise.15em\hbox{$>$}\kern-.85em\lower.35em\hbox{$\sim$}~}}
\def\ls#1{\ifmath{_{\lower1.5pt\hbox{$\scriptstyle #1$}}}}

\def\dslash{\not{\hbox{\kern-2pt $\partial$}}}
\def\Dslash{\not{\hbox{\kern-4pt $D$}}}
\def\Oslash{\not{\hbox{\kern-4pt $O$}}}
\def\Qslash{\not{\hbox{\kern-4pt $Q$}}}
\def\pslash{\not{\hbox{\kern-2.3pt $p$}}}
\def\kslash{\not{\hbox{\kern-2.3pt $k$}}}
\def\qslash{\not{\hbox{\kern-2.3pt $q$}}}

\newcommand{\newc}{\newcommand}
\newc{\qbar}{{\overline q}}
\newc{\TR}{{\rm Tr}}
\newc{\Kahler}{K\"ahler }
\newc{\deltaGS}{\delta_{\rm GS}}

\newc{\Rt}{{\mathbb R}^3}
\newc{\Rf}{{\mathbb R}^4}
\newc{\So}{{\mathbb S}^1}
\newc{\zt}{{\mathbb Z}_2}
\newc{\RtSo}{{\mathbb R}^3\times{\mathbb S}^1}
\newc{\calU}{{\cal U}}
\newc{\calUd}{{\cal U}^\dagger}
\newc{\OG}{\Omega_\Gamma}
\newc{\imat}{{\mathbb I}}
\newc{\sunhc}{SU(N_{\tilde{c}})}

\newcommand{\hpiN}{\tilde{\pi}^{0}}
\newcommand{\heta}{\tilde{\eta}}
\newcommand{\hpiA}{\tilde{\Pi}^A}
\newcommand{\hpiB}{\tilde{\Pi}^B}
\newcommand{\hpi}{\tilde{\pi}}
\newcommand{\hPi}{\tilde{\Pi}}
\newcommand{\fpi}{f_{\tilde{\pi}}}

\newcommand{\thc}{\tilde\theta}
\newcommand{\tqcd}{\theta}
\newcommand{\ggt}{G\tilde{G}}
\newcommand{\hetap}{\tilde{\eta}^\prime}

\begin{document}

\title{Diphotons, New Vacuum Angles, and Strong CP}

\author[a]{Patrick Draper}
\author[b]{and David McKeen}

\affiliation[a]{Department of Physics, University of California, Santa Barbara, CA 93106, USA}
\affiliation[d]{Department of Physics, University of Washington, Seattle, WA 98195, USA}

\emailAdd{pidraper@physics.ucsb.edu}
\emailAdd{dmckeen@uw.edu}

\date{\today}

\abstract{The Standard Model contains a well-understood, natural, spin-0 diphoton resonance: the $\pi^0$. Numerous studies have pointed out that the hint of a new diphoton resonance at 750 GeV could be a pion analog, identified with the pseudo-Nambu-Goldstone boson of a chiral symmetry spontaneously broken by new strong dynamics at the TeV scale. These ``hypercolor" models are generically expected to violate parity through a topological angle $\tilde\theta$. We discuss the physics of $\tilde\theta$ and its impact on the phenomenology of the new sector. 
We also describe some of the theoretical implications of a nonzero $\tilde\theta$.  In particular, $\tilde\theta$ can generate an ${\cal O}(1)$ threshold correction to the QCD vacuum angle $\theta$ near the TeV scale, sharply constraining ultraviolet solutions to the strong CP problem. Alternatively, finding that $\tilde\theta$ is small may be interpreted as evidence in favor of UV solutions to strong CP, particularly those based on spontaneously broken P or CP symmetries.
}


\arxivnumber{16xx.xxxxx}

\preprint{}

\maketitle

\section{Introduction}

The ATLAS and CMS collaborations have both reported modest excesses in diphoton resonance searches near $m_{\gamma\gamma}\simeq 750$ GeV~\cite{CMS:2015dxe,ATLbump}. The appearance of the bump in both experiments in a regime where the background is expected to be featureless is certainly one of the most exciting hints of physics beyond the Standard Model (SM) to date.

An attractive candidate for the diphoton excess at 750 GeV is a neutral pion-like state of a new strongly coupled gauge theory, termed ``hypercolor" in earlier work on vectorlike confinement (VC)~\cite{VC,VC-pheno} (see also related studies~\cite{coloron,coloronLHC,Bai:2010qg,Antipin:2015xia}). The neutral hyperpion $\hpiN$ couples to the QED and QCD topological charge densities through a chiral anomaly, allowing resonant production and decay at the LHC via 
\begin{align}
gg\rightarrow \hpiN\rightarrow\gamma\gamma\;.
\label{channel}
\end{align}
Like the ordinary $\pi^0$ of QCD, due to its composite nature, no scalar mass parameters have to be fine-tuned in order for the hyperpion to remain light.  

A number of groups have studied VC models, new pion-like states, and other pseudo-Nambu-Goldstone boson (PNGB) interpretations of the diphoton excess~\cite{Bellazzini:2015nxw,Low:2015qep,No:2015bsn,Belyaev:2015hgo,Harigaya:2015ezk,Nakai:2015ptz,Franceschini:2015kwy,Molinaro:2015cwg,Matsuzaki:2015che,Bian:2015kjt,Bai:2015nbs,Cline:2015msi,Berthier:2015vbb,Craig:2015lra,Harigaya:2016pnu}. Most studies invoking a new hypercolor sector have been performed in the simplifying limit that the model preserves parity, and in this case the candidate 750 GeV resonance is a pseudoscalar meson. However, a priori, we expect that the new strong dynamics should violate parity through an ${\cal O}(1)$ hypercolor vacuum angle, $\thc$. Here we will study the $\thc$-dependence of physics in the hypercolor sector. (Insofar as models of new strong dynamics are interesting for LHC phenomenology apart from the diphoton excess, this question is also of independent interest, even if the excess is not confirmed by future data.)

In addition to its implications for hypercolor phenomenology, $\thc$ has interesting consequences for the strong CP problem. The same flavor anomalies with QCD that give rise to the production channel~(\ref{channel}) imply that the phases that generate $\thc$ feed directly into the QCD vacuum angle $\tqcd$. Since there is no a priori reason for $\thc$ in particular to be small, the contribution to $\tqcd$ is generically ten orders of magnitude larger than the bound from electric dipole moment measurements~\cite{Baker:2006ts}. 

These new contributions to $\tqcd$ indicate that either there is a new ``hyper-CP problem," or that the strong CP problem must be solved by new physics further in the infrared, such as via the Peccei-Quinn mechanism and its associated axion~\cite{Peccei:1977hh,Peccei:1977ur,Weinberg:1977ma,Wilczek:1977pj}. Thus, $\tqcd$ is a discriminator between solutions to strong CP: the observation of a large $\thc$ would disfavor ultraviolet solutions, while bounding $\thc$ to be small would lend support to models where both $\tqcd$ and $\thc$ are suppressed by the same UV mechanism.

In the case $\thc\sim1$, there is a direct analogy with the electroweak hierarchy problem. The knowledge of the existence of high energy scales like $M_p$, the scale of neutrino masses, and others, through their quantum corrections to the electroweak scale,  tells us that the hierarchy problem is real and must be solved by dynamics around or below those scales. Likewise, a detection of a large CP-violating phase in a new sector at the LHC may indicate that strong CP is {\it not} solved through dynamics at still higher scales, but instead takes place in the infrared.

This paper is organized as follows. In Sec.~\ref{VCtheta} we study the chiral Lagrangian and hyperpion phenomenology of a benchmark model in the presence of $\thc$, including the vacuum structure and existence of Dashen phases~\cite{Dashen:1970et}, the spectrum, and the couplings relevant for collider physics. While parity-preserving couplings are the dominant source of the diphoton signal, parity-violating couplings can lead to large decay rates of heavier hyperpions into pairs of lighter hyperpions in some regions of parameter space, providing an interesting observable signature of nonzero $\thc$. In Sec.~\ref{strongcp} we discuss the impact of $\thc$ on $\tqcd$ in ordinary QCD and the manner in which $\thc$ can be viewed as a discriminator between UV and IR solutions to the strong CP problem. In Sec.~\ref{concl} we summarize and conclude.

We note that Ref.~\cite{Harigaya:2016pnu}, which appeared as this paper was being finished, has some overlap with our study.

\section{Vectorlike Confinement and $\thc$}
\label{VCtheta}
\subsection{Generalities}
The ingredients of VC models~\cite{VC,VC-pheno} are similar to those of QCD: a new asymptotically free  gauge group, ``hypercolor," which we take here to be $\sunhc$, and new vectorlike fermions carrying charges under both hypercolor and the other SM gauge groups. The masses of some of the new fermions are assumed to be less than the strong scale of the $\sunhc$, triggering chiral symmetry breaking and the appearance of light pseudo-Nambu-Goldstone ``hyperpions," among other resonances. Because the SM gauge groups are (gauged) subgroups of the approximate flavor symmetries of the hypercolor sector, typically some hyperpions are charged under SM gauge groups, while others are neutral. Some of the neutral states may decay through the anomaly to pairs of SM gauge bosons, analogously to the decay $\pi^0\rightarrow\gamma\gamma$ in ordinary QCD. A neutral state near the bottom of the spectrum with anomaly-induced couplings to QCD (allowing production through gluon fusion) and QED (allowing decay to diphotons) can provide a candidate for the putative resonance at 750 GeV.

In general the hypercolor sector may possess an arbitrary vacuum angle $\thc$ analogous to the QCD vacuum angle $\tqcd$. In the presence of vectorlike fermions, the microscopic Lagrangian contains the terms
\begin{align}
{\cal L}\supset \frac{\tqcd_0 g^2}{16\pi^2} \TR(\ggt)+\frac{\thc_0 \tilde{g}^2}{16\pi^2} \TR(H\tilde{H})-(M_q q\bar q + h.c.) - (M_\psi \psi\bar \psi+h.c.)\;
\end{align}
where $G$ and $H$ are the color and hypercolor field strengths.\footnote{The duals are defined as $\tilde F_{\mu\nu}\equiv \frac{1}{2}\epsilon_{\mu\nu\rho\sigma}F^{\rho\sigma}$, and we normalize group generators as $\TR(T^aT^b)=\frac{1}{2}\delta^{ab}$.} Here $q$ and $\psi$ ($\bar q$ and $\bar \psi$) are left-handed Weyl fermions taken to be in the fundamental (anti-fundamental) representations of color and hypercolor, respectively. In terms of these parameters, the invariant CP-violating parameters are
$\tqcd$ and $\thc$ are defined as
\begin{align}
\tqcd\equiv&~\tqcd_0+{\rm arg~det~}M_q\nonumber\\
\thc\equiv&~\thc_0+{\rm arg~det~}M_\psi\;.
\end{align}
These expressions are easily generalized in models where some of the vectorlike fermions carry both color and hypercolor, as we will use in this paper.

$\thc$ explicitly breaks parity in the hypercolor sector and has several interesting consequences for the phenomenology of the light hyperpions, including modifying the spectrum and generating parity-violating triple-hyperpion couplings. Also, whereas in the parity-conserving limit, anomaly diagrams lead to pseudo-Goldstone couplings to SM gauge bosons in the form $\TR(\hpi \ggt)$, in the presence of $\thc$, there are additional couplings of the form $\TR(\hpi GG)$. 

In the next subsection we discuss these features concretely in a specific VC benchmark model.

\subsection{Benchmark Model}
Perhaps the simplest model that can accommodate the diphoton excess was studied in~\cite{Nakai:2015ptz}. The model contains two vectorlike fermions, one hypercolor-fundamental QCD-singlet, and the other bifundamental under QCD and hypercolor. Both carry ordinary hypercharge, and neither are charged under $SU(2)_L$. The angle $\thc$ was set to zero in~\cite{Nakai:2015ptz}. 

Let us briefly recall the sense in which this model is ``simplest." The easiest way to resonantly produce a neutral pion-like field is through gluon fusion, and an anomaly-driven coupling to $\ggt$ can be generated if some of the new fermions are colored. Likewise the decay to diphotons may proceed through an anomaly with QED. A model with just one color-triplet fermion produces a color octet hyperpion and a heavy hyper-$\eta^\prime$ ($\hetap$), but no neutral hyperpion. (The $\hetap$ is not a good candidate for the diphoton excess because in that case $\tilde{\Lambda}$ is also of order 750 GeV, and the color octets are too light.) We could add a second color-triplet fermion, in which case the lightest degree of freedom is a neutral hyperpion. However, in this case the QCD anomaly is carried entirely by the $\hetap$, and the light state has to mix with it in order to be resonantly produced through gluon fusion. This mixing is possible, but incalculable in chiral perturbation theory for small $N_{\tilde{c}}$, and furthermore pushes some of the quark masses to rather large values, since the mixing is suppressed in the chiral expansion~\cite{Craig:2015lra}. In the model of~\cite{Nakai:2015ptz}, with one singlet and one triplet vectorlike fermion, there is a light neutral hyperpion $\hpiN$ and it naturally possesses an unsuppressed anomaly coupling to $\ggt$.

This simple model would be sufficient to exhibit the physics of $\thc$ we wish to discuss, including a $\thc$-dependent mass for the light state, parity-violating couplings and decays, and an ${\cal O}(1)$ contribution to $\tqcd$. However, the type of parity-violating decays we will consider proceeds in this model through $\hetap\rightarrow\hpiN\hpiN$, so the relevant coupling is incalculable in chiral perturbation theory. Phenomenologically this is not a problem, but for analytical purposes it is more convenient to discuss a benchmark model with one additional hypercolor-fundamental QCD-singlet flavor. This model contains another light neutral hyperpion, $\heta$, which has calculable parity-violating couplings that permit decays to $\hpiN\hpiN$.

\begin{table}[t!]
\begin{center}
\begin{tabular}{c|c|c|c|c}
  & $\sunhc$ & $SU(3)_{c}$ & $SU(2)_{L}$ & $U(1)_{Y}$  \\
\hline
$\psi_{1}$ & $\Box$ & ${\bf 1}$  & ${\bf 1}$ & $1$  \\
$\bar\psi_{1}$ & $\overline{\Box}$ & ${\bf 1}$ & ${\bf 1}$ & $-1$\\
$\psi_{2}$ & $\Box$ & ${\bf 1}$    & ${\bf 1}$ & $1$  \\
$\bar\psi_{2}$ & $\overline{\Box}$ & ${\bf 1}$ & ${\bf 1}$ & $-1$\\
$\psi_{3}$ & $\Box$ & $\Box$    & ${\bf 1}$ & $-1/3$  \\
$\bar\psi_{3}$ & $\overline{\Box}$ & $\overline{\Box}$ & ${\bf 1}$ & $1/3$
\end{tabular}
\end{center}
\caption{Charge assignments in a simple benchmark VC model.}
\label{tab:charges}
\end{table}

The elementary fields of our benchmark model and their charges are summarized in Table~\ref{tab:charges}. The approximate flavor group of the model is $SU(5)_V\times SU(5)_A$. The axial symmetries are spontaneously broken by chiral condensates $\langle \bar\psi\psi\rangle\sim4\pi f_\pi^3$, and we parametrize the resulting hyperpion Goldstone fields $\hPi$ as
\begin{align}
\Sigma(x)\equiv e^{2i\hPi(x)\cdot T/\fpi}\;.
\label{eq:sigmadef}
\end{align}
where the $T$ generate $SU(5)_A$.

Ordinary color corresponds to gauging the $SU(5)$ generators
\begin{align}
T^a_8\equiv \frac{1}{2}\left(\begin{array}{ccc}
0 & 0 & 0_{1\times 3} \\
0 & 0 & 0_{1\times 3} \\
0_{3\times 1} & 0_{3\times 1} & \lambda^a
\end{array}\right)\;,
\end{align}
under which the 24 hyperpion fields decompose into one color octet, living in block-diagonal elements of $\hPi$; two complex color triplets, living in off-diagonal components of $\hPi$; and four color singlets, two living in diagonal and two living in off-diagonal elements of $\hPi$.\footnote{Gauging subgroups of the vector flavor symmetry has two other important effects. First, the gauging explicitly breaks some of the spontaneously broken axial symmetries, leading to 1-loop masses for the charged hyperpions. Colored hyperpions thus obtain masses that are typically an order-1 factor below the cutoff. Second, the gauging breaks most of the ungauged elements of the vector flavor group, since a general element mixes gauged with ungauged generators. However, some generators may accidentally commute with the gauged elements. In the benchmark model, the two ungauged Cartan elements of $SU(5)_V$ commute with the $SU(3)_c$ generators, yielding an accidental $U(1)^{2}$ ``species symmetry" that is preserved at the renormalizable level~\cite{VC,VC-pheno}. The off-diagonal hyperpions transform under this species symmetry, and the lightest in each species is stable unless higher dimension operators are added that explicitly break the symmetries (alternatively, if the lightest state is neutral, it may provide a DM candidate~\cite{Craig:2015lra}). The hypercharges in the benchmark model here are chosen to allow the triplets to decay through dimension-6 operators of the sort discussed in~\cite{Nakai:2015ptz}. For further discussion of the complications and phenomenology associated with species symmetry, see~\cite{VC,VC-pheno,Nakai:2015ptz}. Since the $\thc$-dependent physics we will study can be illustrated with neutral diagonal hyperpions, we will not need to consider species symmetry, its breaking, or the charged hyperpion states further.}

For our purposes, we can restrict our attention entirely to the two diagonal singlet hyperpions. 
These we refer to as $\hpiA$ and $\hpiB$, corresponding to the axial $SU(5)$ generators
\begin{align}
T_{A}=\frac{1}{2}\left(\begin{array}{ccc}
1 &0&0_{1\times 3} \\
0 & -1&0_{1\times 3}\\
0_{3\times 1}&0_{3\times 1}&0_{3\times 3}
\end{array}\right)\;\;\;\;T_{B}=\frac{1}{\sqrt{15}}\left(\begin{array}{ccc}
-\frac{3}{2} &0&0_{1\times 3} \\
0 & -\frac{3}{2}&0_{1\times 3}\\
0_{3\times 1}&0_{3\times 1}&1_{3\times 3}
\end{array}\right)\;.
\end{align}
The axial transformations generated by $T_{B}$ are anomalous under QCD. Also, the axial transformations generated by
\begin{align}
T_{\heta^\prime}=\frac{1}{\sqrt{10}}\left(\begin{array}{ccc}
1 &0&0_{1\times 3} \\
0 & 1&0_{1\times 3}\\
0_{3\times 1}&0_{3\times 1}&1_{3\times 3}
\end{array}\right)
\end{align}
are anomalous with both QCD and hypercolor.

The leading $\thc$-dependent terms in the chiral lagrangian are generated by the hyperpion mass terms,
\begin{align}
{\cal L}\supset
\mu\frac{\fpi^2}{2}\TR[\Sigma^\dagger M+M^\dagger\Sigma]\;.
\label{eq:massterm}
\end{align}
$\mu$ is a scale parameter expected to be of order $\tilde{\Lambda}\sim 4\pi \fpi/\sqrt{N_{\tilde{c}}}$. 
Without using the axial transformations anomalous under color, the mass matrix in our benchmark model may be brought into the form
\begin{align}
M_0=\left(\begin{array}{ccc}
M_1 e^{i\phi_1} & 0 & 0_{1\times 3} \\
0 & M_2 e^{i\phi_2} & 0_{1\times 3} \\
0_{3\times 1} & 0_{3\times 1} & M_3 e^{i\phi_3/3}\times1_{3\times3}
\end{array}\right)\;.
\label{eq:M0}
\end{align}
where $M_{1,2,3}$ are real. Each sub-block corresponds to fields that form a representation under a gauged subgroup of the diagonal flavor symmetry. Because the axial transformation generated by $T_{B}-\sqrt{2/3}T_{\heta^\prime}$ is anomalous with hypercolor, but not color, we may assume that the hypercolor vacuum angle has already been moved to reside entirely in ${\rm arg~det~}M_0$ without shifting the QCD $\tqcd$ term. Thus
\begin{align}
\thc\equiv\phi_1+\phi_2+\phi_3\;.
\end{align}
Subsequently, using $T_{B}$ and $T_{A}$ transformations, we may take the mass matrix into the form
\begin{align}
M=\left(\begin{array}{ccc}
M_1e^{i\thc/2} & 0 & 0_{1\times 3} \\
0 & M_2e^{i\thc/2} & 0_{1\times 3} \\
0_{3\times 1} & 0_{3\times 1} & M_3\times1_{3\times3}
\end{array}\right)\;
\label{eq:M1}
\end{align}
which we will use to obtain the hyperpion Lagrangian terms in Eq.~(\ref{eq:massterm}). Since the $T_{B}$ transformation that brings the matrix~(\ref{eq:M0}) into the form~(\ref{eq:M1}) is anomalous with QCD, it shifts $\tqcd$ by an amount
\begin{align}
(\Delta\tqcd)_1=\frac{N_{\tilde{c}}}{3}\phi_3\;.
\label{eq:dtqcd1}
\end{align}
Another way to say it is we have a new contribution to ${\rm arg~det~}M_q$, where $M_q$ is the colored fermion mass matrix, coming from $\psi_3$. Eq.~(\ref{eq:dtqcd1}) is one of two contributions to $\Delta\tqcd$ from the hypercolor sector. We discuss the second and their implications further in Sec.~\ref{strongcp}.

\subsection{Neutral Sector Phenomenology for $M_{1,2}\ll M_3$}
\label{sec:analyt}
As mentioned above, we can use the two neutral diagonal hyperpions $\hpiA$ and $\hpiB$ to illustrate various effects of $\thc$. In this section we will study the physics of the neutral hyperpions in simplifying limits amenable to analytic treatment, in particular the ``QCD-like" limit
\begin{align}
M_1,M_2\ll M_3\;.
\label{eq:qcdlim}
\end{align}
In the next section we perform precise numerical analysis on a broader range of parameter space, but the analysis here in the limit~(\ref{eq:qcdlim}) will help us understand qualitative features.

The potential for $\hpiA$ and $\hpiB$ arising from~(\ref{eq:massterm}) with mass matrix~(\ref{eq:M1}) is given by
\begin{align}
V(\hpiA,\hpiB)=
-\fpi^2&\mu \bigg[M_1\cos\left(\frac{\thc}{2}-\frac{\hpiA}{\fpi}+\sqrt{\frac{3}{5}}\frac{\hpiB}{\fpi}\right)\nonumber\\
&+M_2\cos\left(\frac{\thc}{2}+\frac{\hpiA}{\fpi}+\sqrt{\frac{3}{5}}\frac{\hpiB}{\fpi}\right)
+3M_3\cos\left(\frac{2\hpiB}{\sqrt{15}\fpi}\right)\bigg]\;.
\label{eq:Vpieta}
\end{align}
For $\thc\neq 0$, this potential is minimized for nonzero $\hpiA$, $\hpiB$. 

To analyze the vacuum structure, we may eliminate $\hpiA$ with its equation of motion,
\begin{align}
\tan\left(\frac{\hpiA}{\fpi}\right)=\left(\frac{M_1-M_2}{M_1+M_2}\right)\tan\left(\frac{\thc}{2}+\sqrt{\frac{3}{5}}\frac{\hpiB}{\fpi}\right)\;,
\end{align}
after which the equation for $\hpiB$ reduces to
\begin{align}
M_3\sin\left(\frac{2}{\sqrt{15}}\frac{\hpiB}{\fpi}\right)=\mp\frac{M_1M_2\sin\left(\thc+2\sqrt{\frac{3}{5}}\frac{\hpiB}{\fpi}\right)}{\sqrt{M_1^2+M_2^2+2M_1M_2\cos\left(\thc+2\sqrt{\frac{3}{5}}\frac{\hpiB}{\fpi}\right)}}\;,
\label{eq:hetamin}
\end{align}
with the upper sign (-) corresponding to the solutions with the lowest energy. The vacuum structure reflected by Eq.~(\ref{eq:hetamin}) is a highly nontrivial function of the input parameters.  In certain regimes of the hyperquark masses, (\ref{eq:hetamin}) has multiple solutions, analogous to the Dashen phenomenon of QCD at $\tqcd=\pi$~\cite{Dashen:1970et} and Witten's generalization to other values of $\tqcd$~\cite{Witten:1980sp}.  In QCD, the global minimum of $V(\pi^0,\eta)$ is a non-analytic function of $\tqcd$ when multiple vacua exist, with the energies of different vacua crossing at $\tqcd=\pi$~\cite{Witten:1980sp}.

 \begin{figure}[t]
  \centering
 \includegraphics[height=0.40\textwidth]{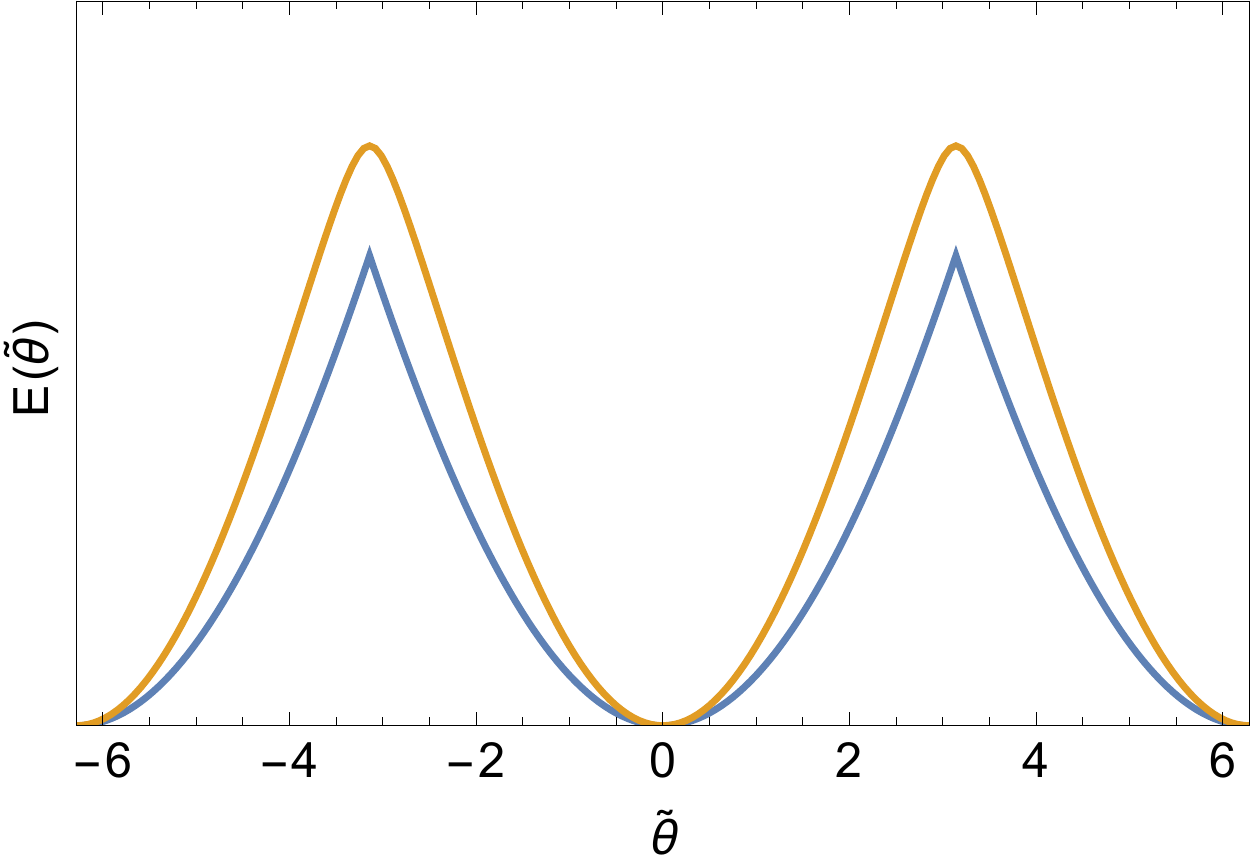}
  \caption{The behavior of the vacuum energy as a function of $\thc$. In the lower curve, we take $M_1=200$ GeV, $M_2=300$ GeV, and $M_3=500$ GeV, so that Eq.~(\ref{eq:dashenphase}) is satisfied. Correspondingly the potential ~(\ref{eq:Vpieta}) exhibits both global and local minima, and the vacuum energy is cuspy across $\thc=\pi$, where local and global minima exchange roles. In the upper curve, we take $M_1=200$ GeV, $M_2=300$ GeV, and $M_3=5$ TeV, so that Eq.~(\ref{eq:dashenphase}) is not satisfied, and the vacuum energy is a smooth function of $\thc$.}
  \label{fig:dashenphase}
\end{figure}

Structure similar to this ``multi-branched" behavior may also arise in our hypercolor theory, and is easiest to exhibit at $\thc=\pi$. Here Eq.~(\ref{eq:hetamin}) always has at least one solution,
\begin{align}
\hpiB/\fpi=0\;.
\end{align}
Factoring out this root, we may rearrange the $\hpiB$ equation to read
\begin{align}
M_1^2M_2^2(3-4\sin^2\alpha)^2&=M_3^2(M_1^2+M_2^2+2M_1M_2\sqrt{1-\sin^2\alpha}(-1+4\sin^2\alpha))\;,
\label{eq:hetamin2}
\end{align}
where we have defined $\alpha\equiv 2\hpiB/\sqrt{15}\fpi$. In the simplifying regime~(\ref{eq:qcdlim}), in order for Eq.~(\ref{eq:hetamin2}) to exhibit new solutions, we evidently require that the splitting between $M_1$ and $M_2$ is small compared to $M_{1,2}$, in which case new solutions may appear at small $\alpha$. In the regime
\begin{align}
\frac{M_1}{M_3}\sim\frac{|M_1-M_2|}{M_1}\ll 1\;,
\end{align}
a new pair of $\hpiB/\fpi$ vacua exists when 
\begin{align}
3M_1M_2>M_3|M_1-M_2|\;.
\label{eq:dashenphase}
\end{align}
Condition~(\ref{eq:dashenphase}) is completely analogous to the Dashen phase condition $m_um_d>m_s|m_u-m_d|$ in QCD at $\tqcd=\pi$, and when the nonzero solutions exist, they are (degenerate) global minima in which parity is spontaneously broken. For other values of $\tqcd$ and $\thc$, the levels are split and there is one global minimum. The global and local minima exchange roles as $\thc$ crosses $\pi$. This behavior is illustrated in Fig.~\ref{fig:dashenphase}, where we plot the energy of the global minimum as a function of $\thc$ for values of the $M_i$ both satisfying and violating Eq.~(\ref{eq:dashenphase}). In the former case, the energy is cuspy at $\thc=\pi+2\pi k$, corresponding to the crossing of branches; in the latter case, the energy is smooth.

For general $\thc\neq\pi$, the phenomenology in which we are interested is qualitatively the same regardless of whether Eq.~(\ref{eq:hetamin}) has exactly one or multiple solutions. Therefore, for simplicity we will now focus on the limit~(\ref{eq:qcdlim}) {\it without} strong degeneracy between $M_1$ and $M_2$.  In this regime there is only one solution, and it is convenient to reanalyze the potential, starting by integrating out the $\hpiB$.
To first order in $M_{1,2}$, the solution for $\hpiB$ is
\begin{align}
\frac{\hpiB}{\fpi}= -\frac{\sqrt{15}}{4}\frac{M_1}{M_3}\sin\left(\frac{\thc}{2}-\frac{\hpiA}{\fpi}\right)-\frac{\sqrt{15}}{4}\frac{M_2}{M_3}\sin\left(\frac{\thc}{2}+\frac{\hpiA}{\fpi}\right)\;,
\label{eq:etasol}
\end{align}
reflecting the fact that mixing between the $\hpiB$ and $\hpiA$ states is controlled by $\frac{M_2-M_1}{M_3}$ in the limit~(\ref{eq:qcdlim}). Eq.~(\ref{eq:etasol})
 generates an effective potential for the light field,
\begin{align}
V(\hpiA)=-f^2\mu \left[M_1\cos\left(\frac{\thc}{2}-\frac{\hpiA}{\fpi}\right)+M_2\cos\left(\frac{\thc}{2}+\frac{\hpiA}{\fpi}\right)\right]\;,
\label{eq:vhpi}
\end{align}
and $V(\hpiA)$ is minimized by
\begin{align}
\tan\left(\frac{\hpiA}{\fpi}\right)=\frac{M_1-M_2}{M_1+M_2}\tan\left(\frac{\thc}{2}\right)\;.
\end{align}
Plugging back into Eq.~(\ref{eq:etasol}), the corresponding vev for the $\hpiB$ field is
\begin{align}
\frac{\hpiB}{\fpi}=-\frac{\sqrt{15}M_1M_2\sin(\thc)}{2M_3\sqrt{M_1^2+M_2^2+2M_1M_2\cos(\thc)}}\;.
\label{eq:veta}
\end{align}

In general, the states $\hpiA$ and $\hpiB$ undergo mass mixing. We will refer to the lighter mass eigenstate as $\hpiN$ and the heavier as $\heta$. Unlike QCD, the states may be heavily mixed. However, in the limit analyzed in this section, the spectrum is insensitive to mixing at first order. The masses are given by
\begin{align}
m^2_{\hpiN}&=\mu \sqrt{M_1^2+M_2^2+2M_1M_2\cos(\thc)}\nonumber\\
m^2_{\heta}&=\frac{4}{5}\mu M_3+\frac{3}{5}m^2_{\hpiN}\;.
\label{eq:pionmasses}
\end{align}
While for $\thc=0$ the $\hpiN$ mass grows with $M_1$ and $M_2$, for $\thc$ of order $\pi$, the mass is controlled by the difference $|M_1-M_2|$.

In the above approximations, the diagonal octet mass 
\begin{align}
m^2_8=2\mu M_3+(a\tilde\Lambda)^2\;,
\end{align}
where $a\simeq 0.3$ parametrizes the effects of loop corrections from QCD. $m^2_8$ is sensitive to $\thc$ only at subleading order in $M_1/M_3$, and then only through the expectation value for $\hpiB$.

A particularly interesting feature of nonzero $\thc$ is the appearance of a large number of parity-violating cubic couplings in the hyperpion potential. Again is it sufficient study the potential~(\ref{eq:Vpieta}) in the limit $M_{1,2}\ll M_3$. To zeroth order, the cubic couplings are:
\begin{align}
V_{\rm cubic}=-\frac{\mu}{\fpi}\frac{\sin(\thc)}{3\sqrt{15}}\frac{M_1M_2}{ \sqrt{M_1^2+M_2^2+2M_1M_2\cos(\thc)}}\left(\heta\heta\heta+9\heta\hpiN\hpiN\right)\;.
\label{eq:cubics}
\end{align}
$V_{\rm cubic}$ allows the parity-violating decay $\heta\rightarrow\hpiN\hpiN$ when kinematically allowed.

At the LHC, the most important couplings for the neutral hyperpions are to the QCD and QED anomalies, which allow production through gluon fusion and decay to diphotons even in the absence of parity violation.
Before mixing, only the $\hpiB$ field couples to the QCD $\ggt$ and QED $F\tilde F$,\footnote{There are also similar couplings to $ZZ$ and $Z\gamma$.}
\begin{align}
{\cal L}&\supset\frac{N_{\tilde{c}}\alpha_s}{2\pi\fpi}~\TR(T_{B} T_8^aT_8^b)~\hpiB~G^a\tilde{G}^b+\frac{N_{\tilde{c}}\alpha}{4\pi\fpi}~\TR(T_{B} QQ)~\hpiB~F\tilde{F}\nonumber\\
&\Rightarrow \frac{N_{\tilde{c}}\alpha_s}{2\pi\fpi}\frac{1}{\sqrt{15}}~\hpiB~\TR(\ggt)-\frac{N_{\tilde{c}}\alpha}{4\pi\fpi}\frac{8}{3\sqrt{15}}~\hpiB~F\tilde{F}\;.
\label{eq:etaGGt}
\end{align}
These couplings leading to the resonant process $pp\rightarrow\heta\rightarrow\gamma\gamma$ at the LHC, offering a discovery mode for $\heta$ when the diphoton branching ratio is unsuppressed.
Mixing induced by~(\ref{eq:etasol}) also generates $\hpiN\ggt$ and $\hpiN F\tilde{F}$ couplings when $\heta$ is integrated out,
\begin{align}
{\cal L}\supset c&\left[\frac{N_{\tilde{c}}\alpha_s}{2\pi\fpi}\frac{1}{\sqrt{15}}~\hpiN~\TR(\ggt)-\frac{N_{\tilde{c}}\alpha}{4\pi\fpi}\frac{8}{3\sqrt{15}}~\hpiN~F\tilde{F}\right]\;,\nonumber\\
c&\equiv\frac{(M_1^2-M_2^2)}{8M_3\sqrt{M_1^2+M_2^2+2M_1M_2\cos(\thc)}}\;
\label{eq:piGGt}
\end{align}

These couplings can be large if either $M_1$ or $M_2$ is not substantially smaller than $M_3$, and lead to the resonant process~(\ref{channel}).

For nonzero $\thc$, we expect the effective theory should also contain other parity violating couplings allowed by the symmetries. For example, for small $\thc$, we expect a coupling of the form
\begin{align}
{\cal L}\supset c^\prime\cdot\frac{\alpha_s(\tilde\Lambda)}{2\pi}\frac{M\thc}{\fpi\tilde\Lambda}\hpiN \TR(GG)\sim c\cdot\alpha_s(\tilde\Lambda)\frac{M\thc}{\tilde\Lambda^2}\hpiN \TR(GG)\;,
\label{eq:piGG}
\end{align}
where $G$ is the QCD field strength and $M$ is a characteristic hyperquark mass. Unlike the anomaly-generated coupling $\hpiN \TR(G\tilde G)$, we cannot compute $c^\prime$ in Eq.~(\ref{eq:piGG}). Relative to the anomaly coupling it is also chirally suppressed, so~(\ref{eq:piGGt}) still plays the dominant role in $\hpiN$ resonant production.

\subsection{Numerical Analysis and Diphoton Rates}

In the previous section we discussed the impact of $\thc$ and the quark mass parameters on the properties of the benchmark model in a special limit amenable to analytic treatment. 
Here we illustrate some of these features quantitatively and extend the analysis numerically to more general parameter regimes. (However, we postpone one phenomenological question - that of the parity-violating decays $\heta\rightarrow\hpiN\hpiN$ - for Sec.~\ref{sec:pvd}.)

\begin{figure}[t]
  \centering
 \includegraphics[height=0.4\textwidth]{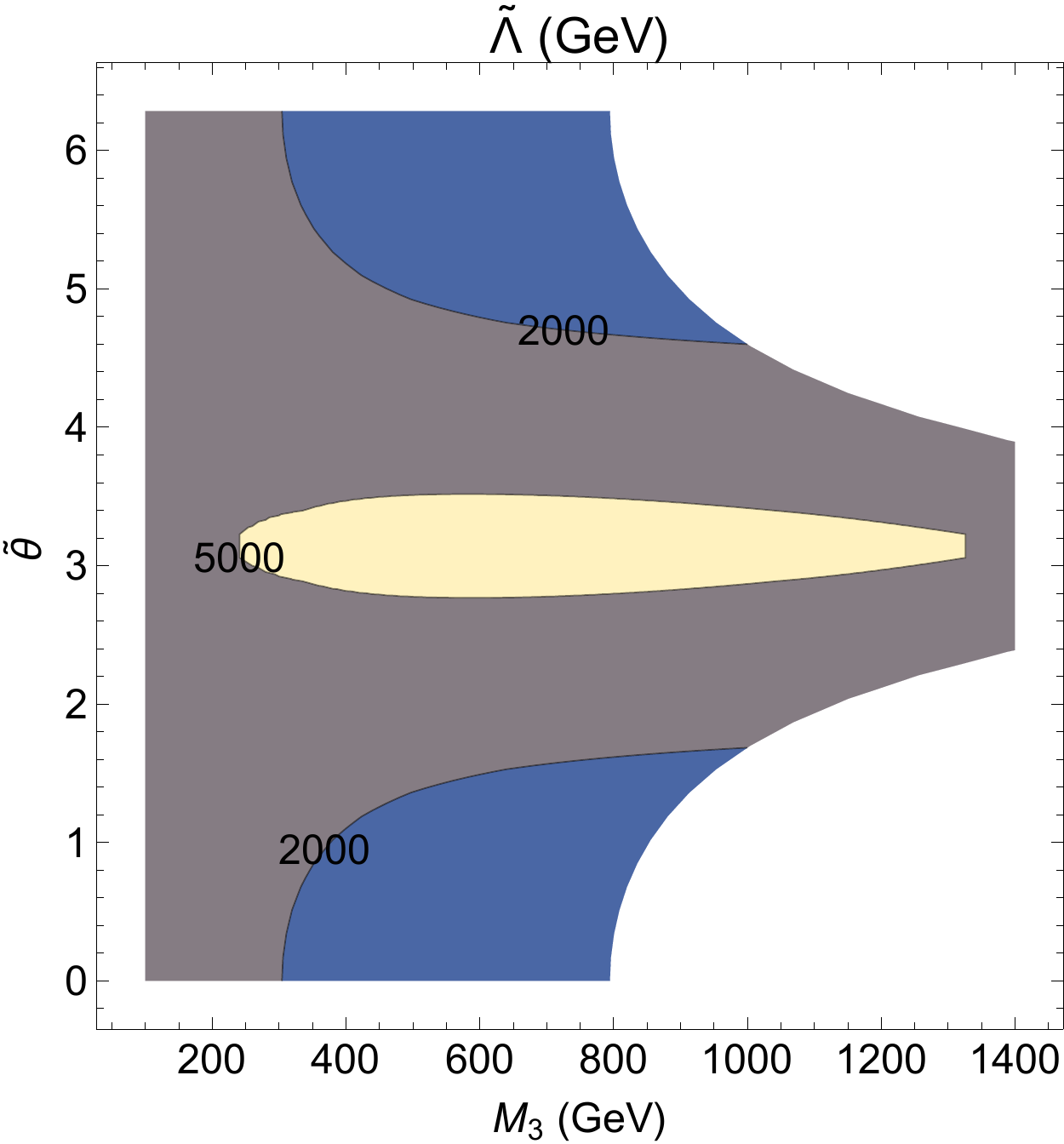}\;\;\;\;\;\;\;
  \includegraphics[height=0.4\textwidth]{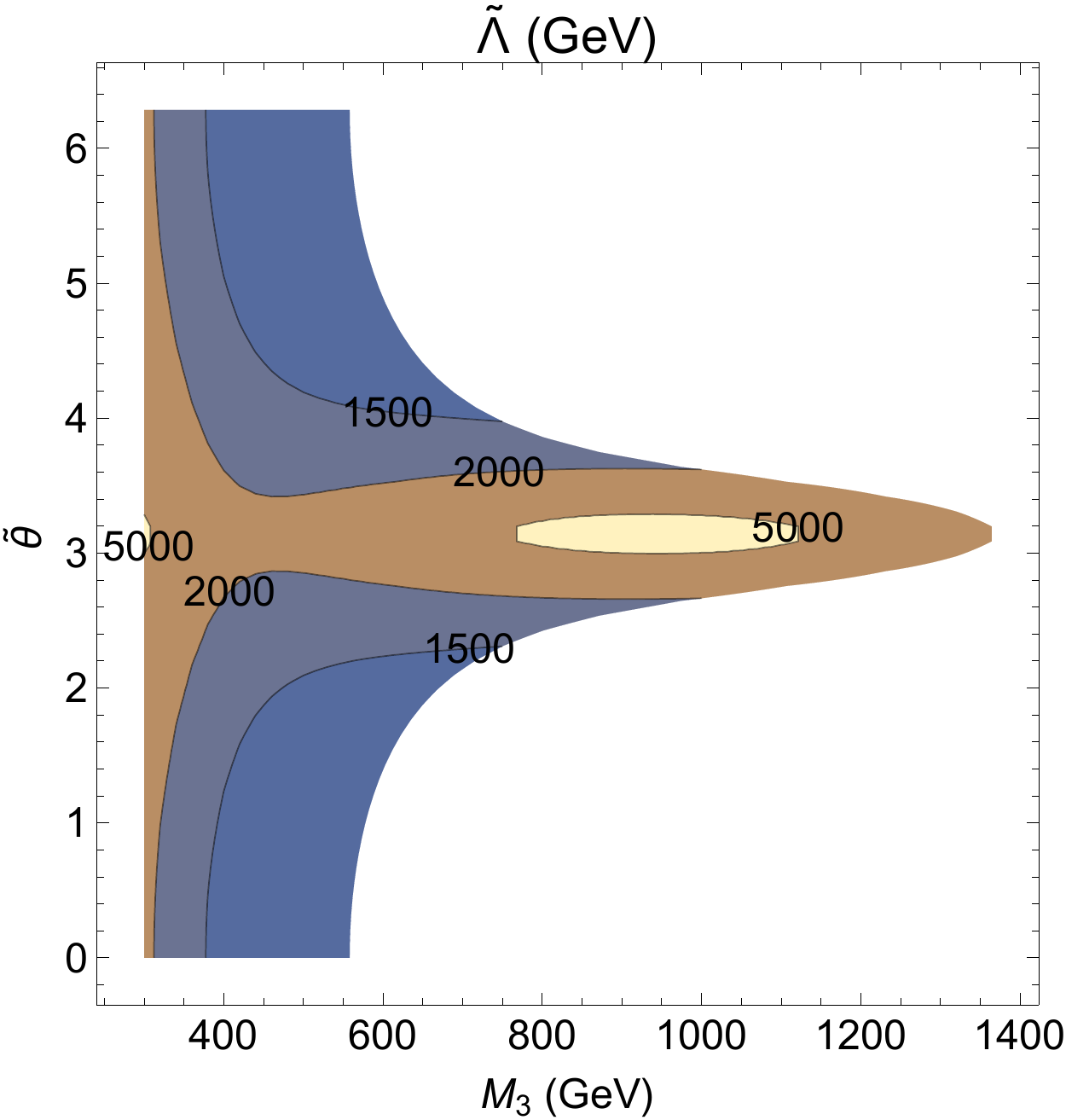}\\
~\\
   \includegraphics[height=0.4\textwidth]{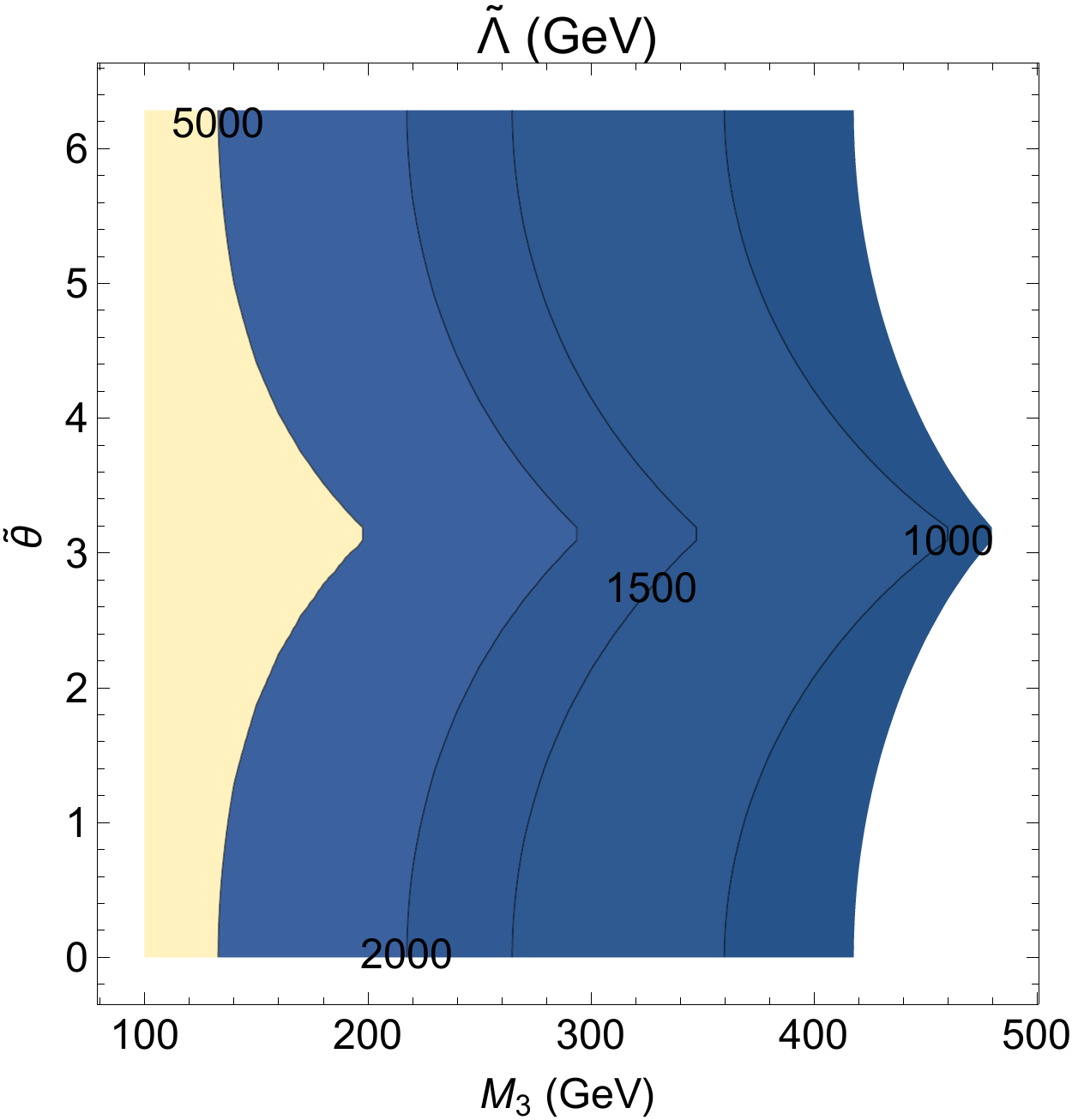}\;\;\;\;\;\;\;
  \includegraphics[height=0.4\textwidth]{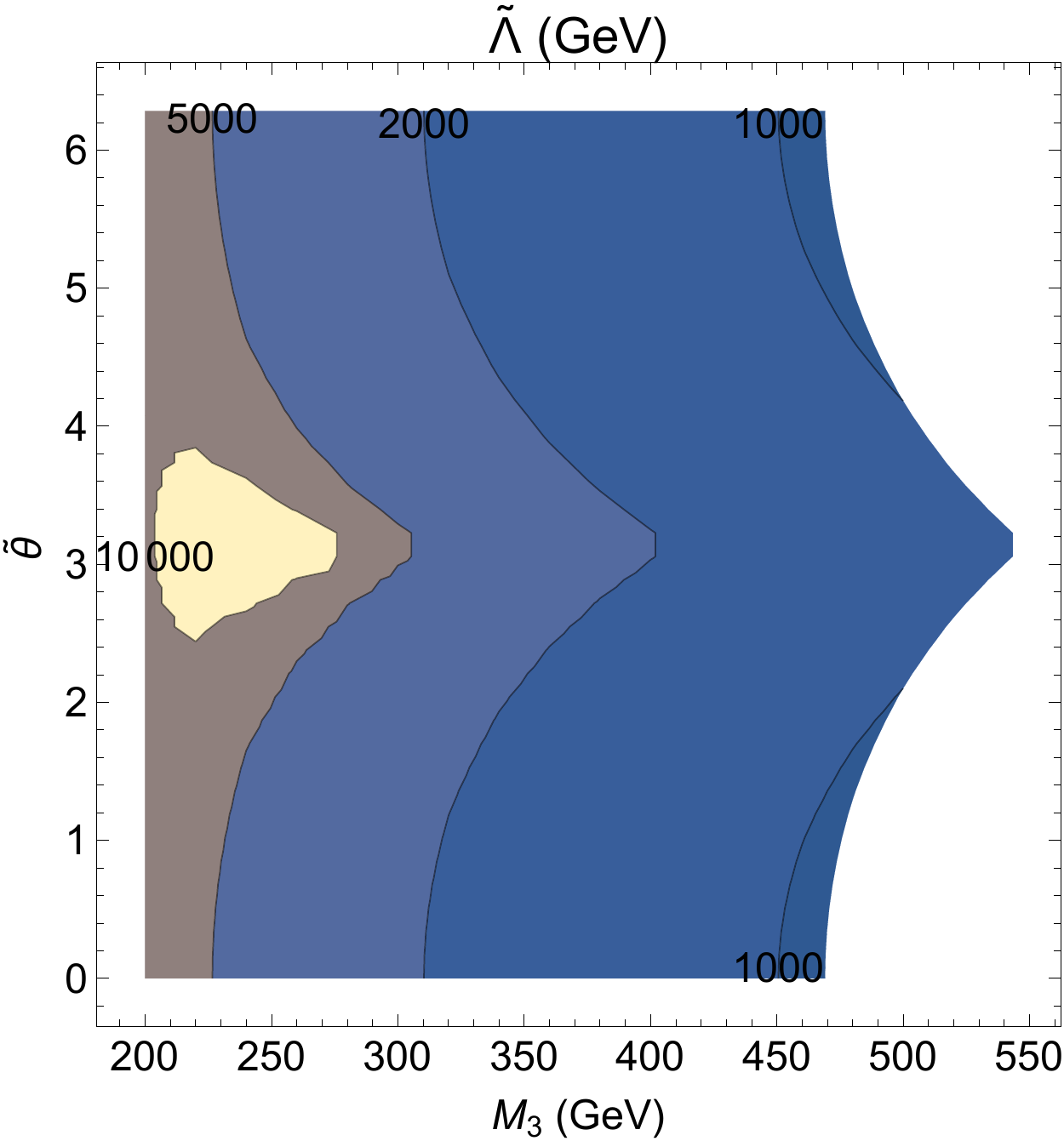}
  \caption{Benchmark model results for the cutoff $\tilde\Lambda$ for different ranges of hyperfermion masses. In all cases $\fpi$ (and consequently $\tilde\Lambda$) is fixed so that the lightest neutral state has mass $m_{\hpiN}=750$ GeV. We plot on the ($M_3,\thc$) plane, and plots are clipped where any of the masses $M_i$ exceeds half the cutoff $\tilde\Lambda$. First row: $M_1=100$ GeV, $M_2=300$ GeV (left); $M_1=200$ GeV, $M_3-M_2=200$ GeV (right). Second row: $M_1=400$ GeV, $M_3-M_2=100$ GeV (left); $M_2-M_1=M_3-M_2=100$ GeV (right). }
  \label{fig:cutoff}
\end{figure}

\begin{figure}[t]
  \centering
 \includegraphics[height=0.4\textwidth]{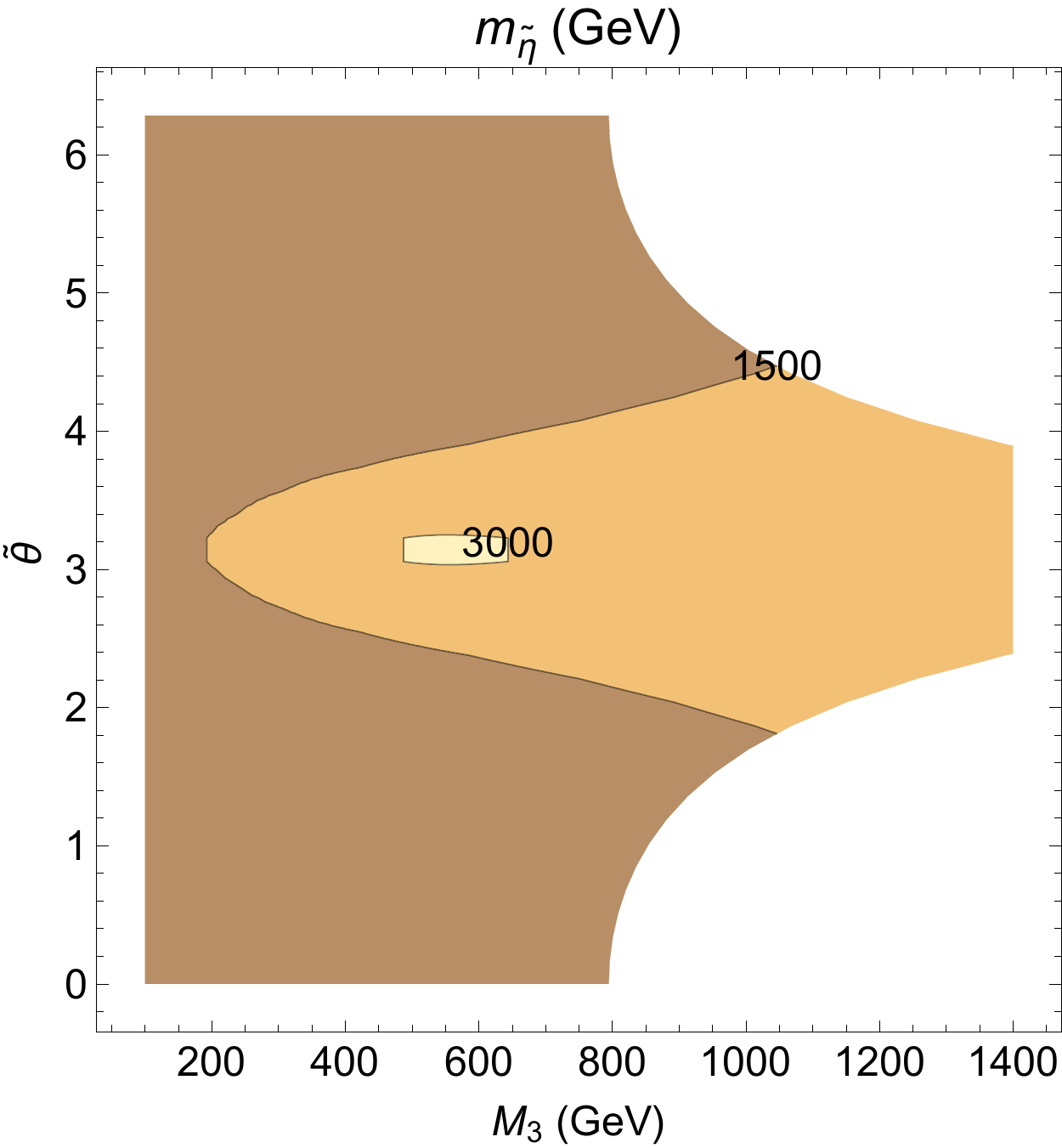}\;\;\;\;\;\;\;
  \includegraphics[height=0.4\textwidth]{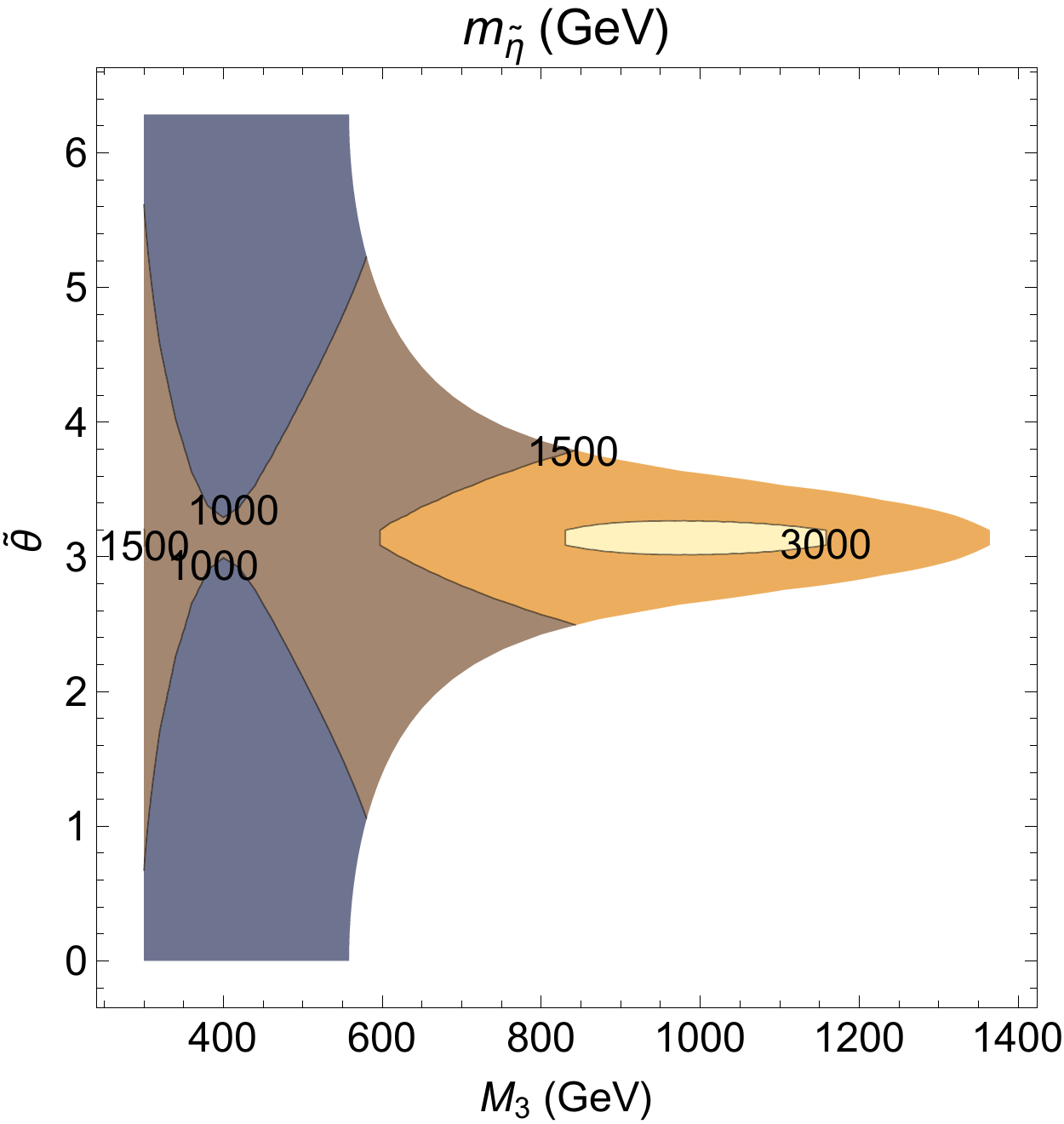}\\
~\\
   \includegraphics[height=0.4\textwidth]{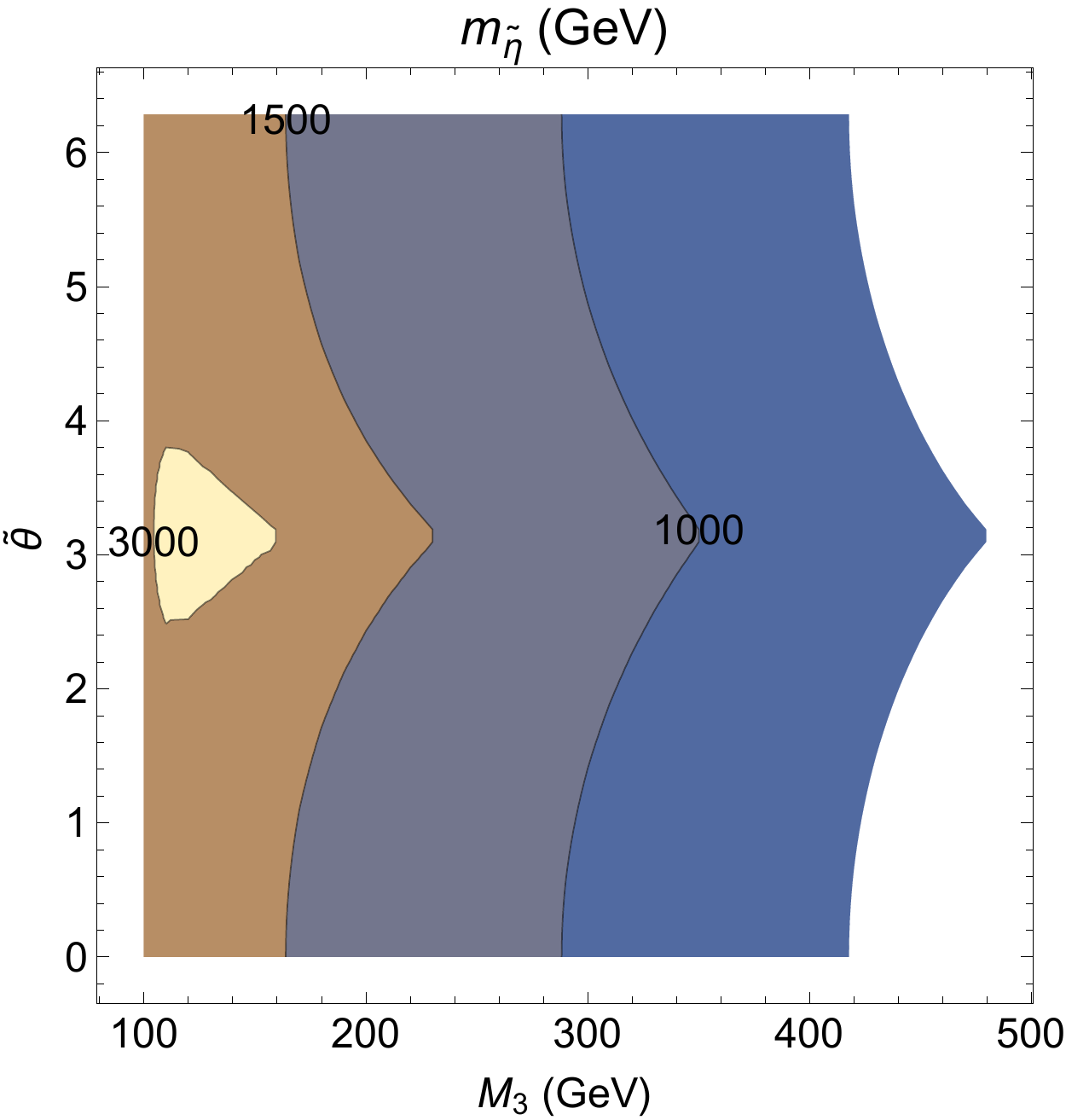}\;\;\;\;\;\;\;
  \includegraphics[height=0.4\textwidth]{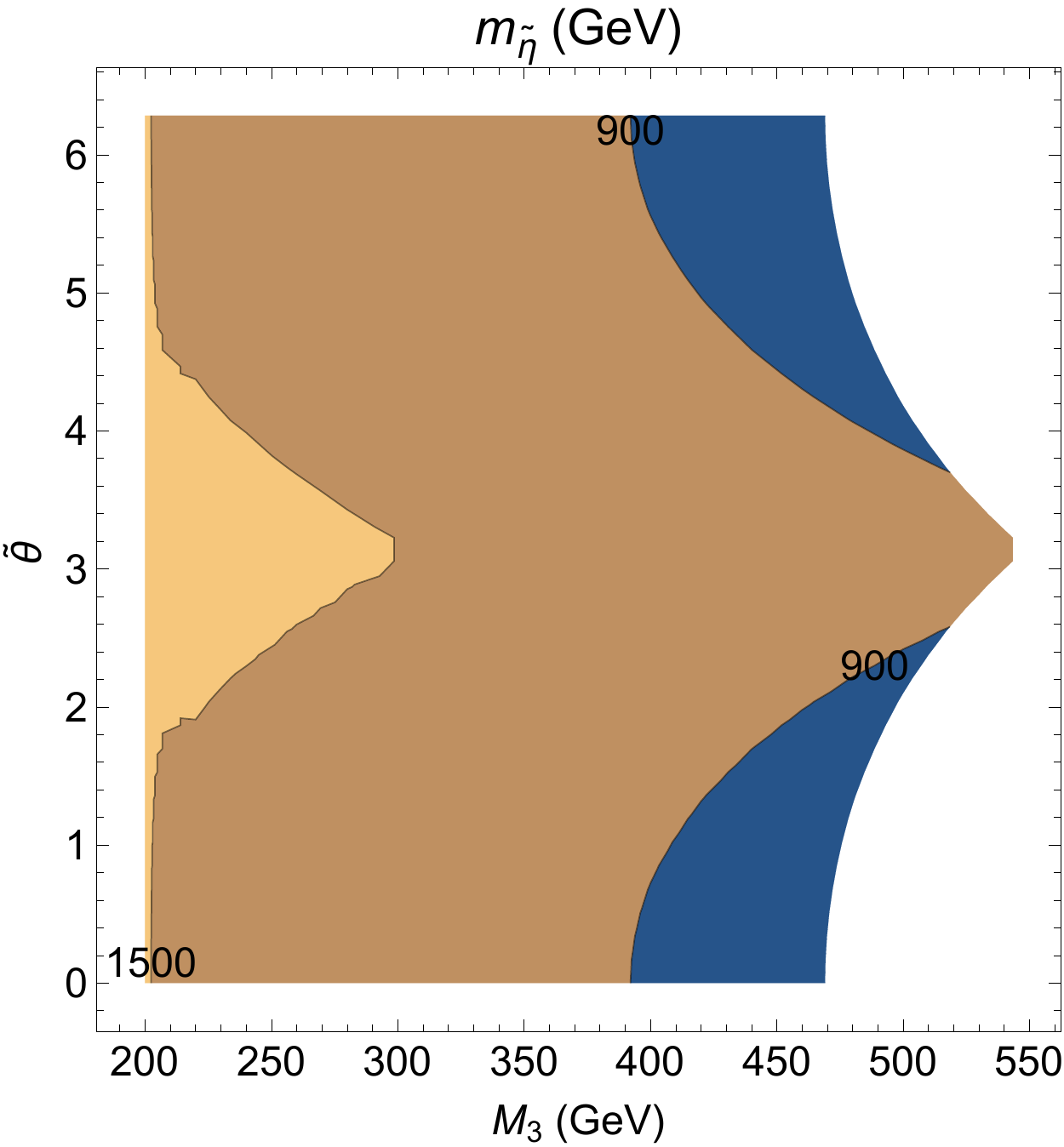}
  \caption{Benchmark model results for $m_{\heta}$. In all cases $\fpi$ (and consequently $\tilde\Lambda$) is fixed so that the lightest neutral state has mass $m_{\hpiN}=750$ GeV. We plot on the ($M_3,\thc$) plane, and plots are clipped where any of the masses $M_i$ exceeds half the cutoff $\tilde\Lambda$. First row: $M_1=100$ GeV, $M_2=300$ GeV (left); $M_1=200$ GeV, $M_3-M_2=200$ GeV (right). Second row: $M_1=400$ GeV, $M_3-M_2=100$ GeV (left); $M_2-M_1=M_3-M_2=100$ GeV (right). }
  \label{fig:meta}
\end{figure}

Fixing $\mu=\tilde{\Lambda}$ (which can be taken as a definition of the overall scale of the hyperfermion masses) and choosing $\tilde N=3$, there are five free parameters in the benchmark model, given by $\fpi$, the three $M_i$, and $\thc$.  We fix $\fpi$ by requiring that the lightest neutral state has mass $750$ GeV for each value of the hyperfermion masses and $\thc$, and we analyze the potential~(\ref{eq:Vpieta}) in four different parameter scenarios.
\begin{enumerate}
\item{{\bf Scenario 1:} $M_1=100$ GeV, $M_2=300$ GeV. As $M_3$ becomes larger than $M_{1,2}$, we approach the regime analyzed in the previous section, with a ``mostly-$\hpiA$" $\hpiN$ state and a ``mostly-$\hpiB$" $\heta$ state.}
\item{{\bf Scenario 2:}  $M_1=200$ GeV, $M_3-M_2=200$ GeV, focusing on $M_{2,3}>M_1$. Here the states are well-mixed, and as two of the masses become large the cutoff must come down to maintain $m_{\hpiN}=750$ GeV.}
\item{{\bf Scenario 3:} $M_1=400$ GeV, $M_3-M_2=100$ GeV, focusing on $M_{2,3}<M_1$. Again the states are well -mixed, but small masses imply that the cutoff is large.}
\item{{\bf Scenario 4:} $M_2-M_1=M_3-M_2=100$ GeV. The states are well-mixed and quasi-degenerate, and we vary the overall mass scale.}
\end{enumerate}

Much of the phenomenology in each of the four scenarios is governed by the cutoff $\tilde{\Lambda}=4\pi\fpi/\sqrt{\tilde N}$, which in turn is fixed by the requirement $m_{\hpiN}=750$ GeV. In Fig.~\ref{fig:cutoff} we plot
$\tilde{\Lambda}$. In most cases the cutoff increases as $\thc$ approaches $\pi$, reflecting the fact that terms in $m_{\hpiN}$ begin to cancel against each other when $\cos(\thc)<0$. This behavior is evident in Eq.~(\ref{eq:pionmasses}) in the regime $M_1\sim M_2\ll M_3$. In Scenario 2, the cancellation is particularly efficient for a range of masses around $M_3\sim 1 $ TeV.
In Fig.~\ref{fig:meta} we plot the mass of the heavier neutral Goldstone, which largely tracks the features of $\tilde\Lambda$. Octet masses (not shown) exhibit similar behavior and are of the order 1-2 TeV in all scenarios.

\begin{figure}[t]
  \centering
 \includegraphics[height=0.4\textwidth]{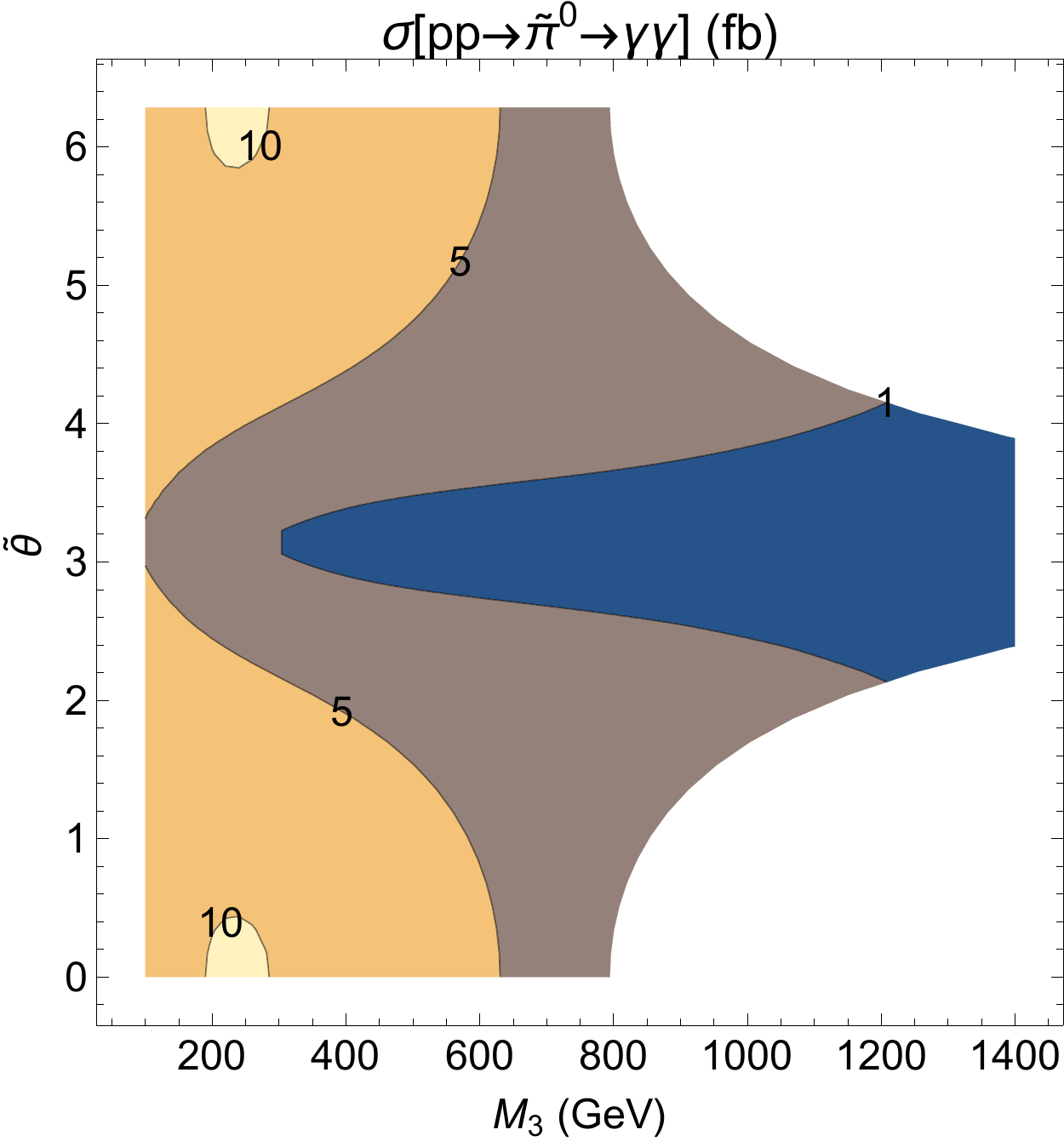}\;\;\;\;\;\;\;
  \includegraphics[height=0.4\textwidth]{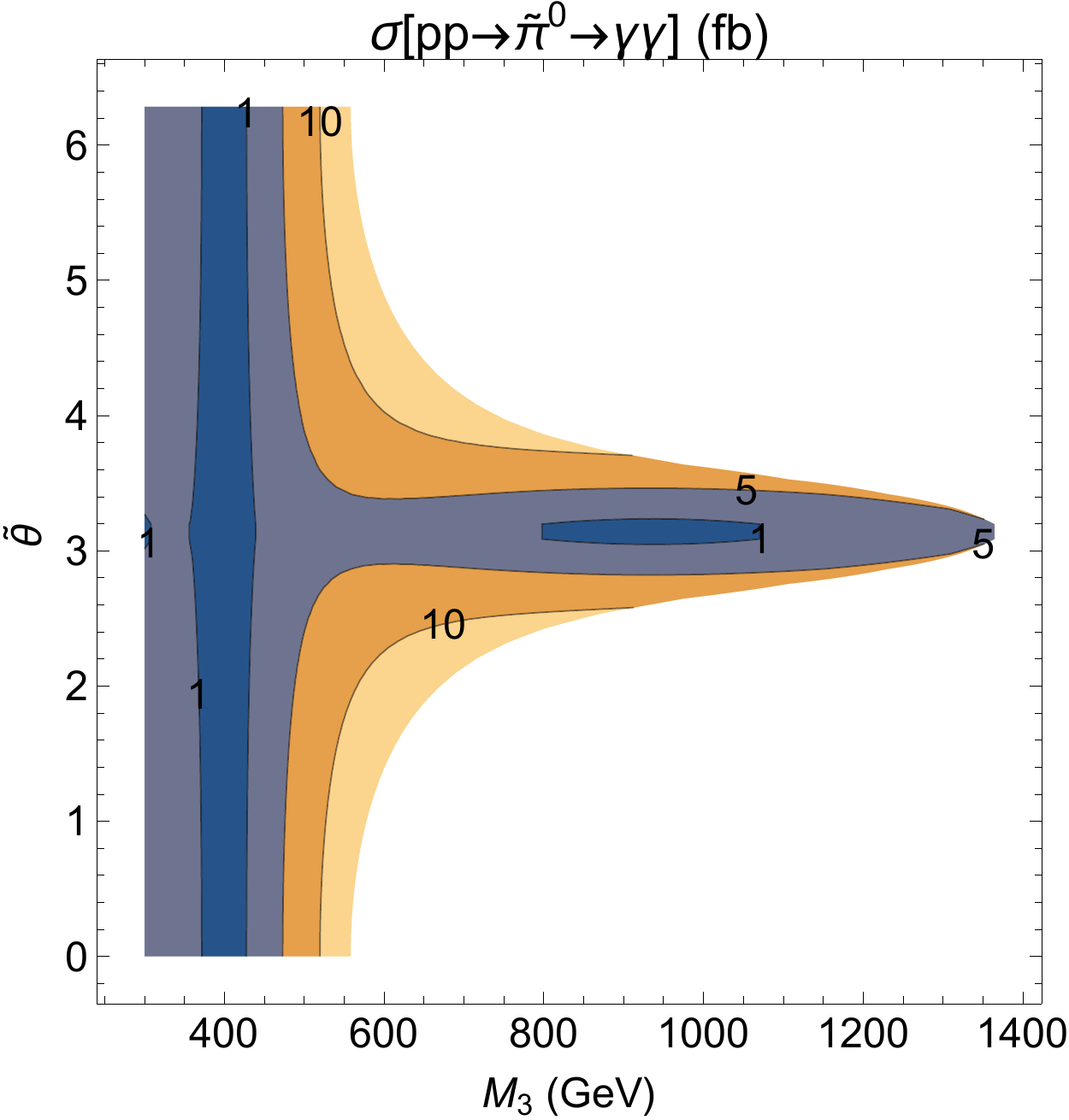}\\
~\\
   \includegraphics[height=0.4\textwidth]{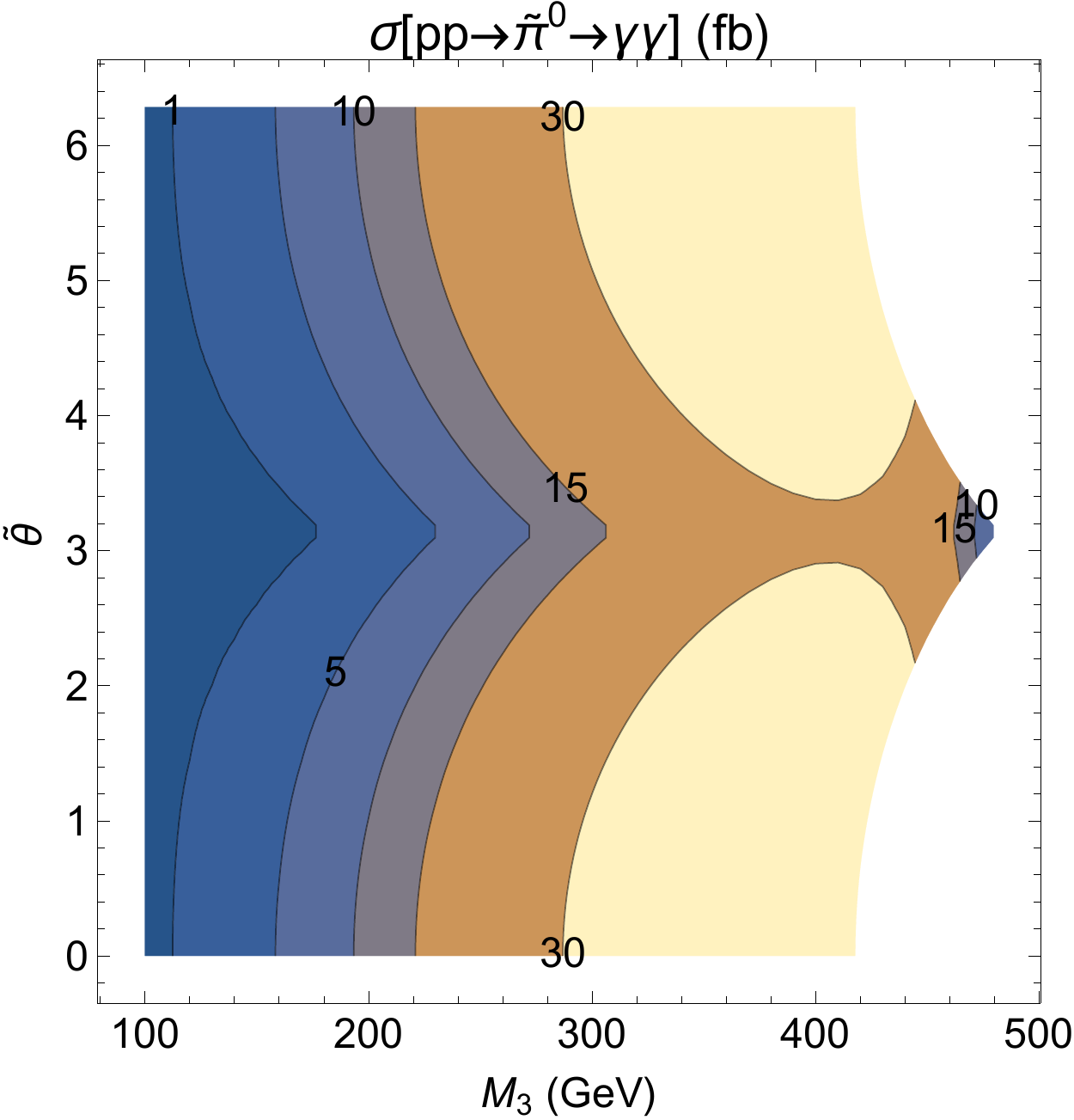}\;\;\;\;\;\;\;
  \includegraphics[height=0.4\textwidth]{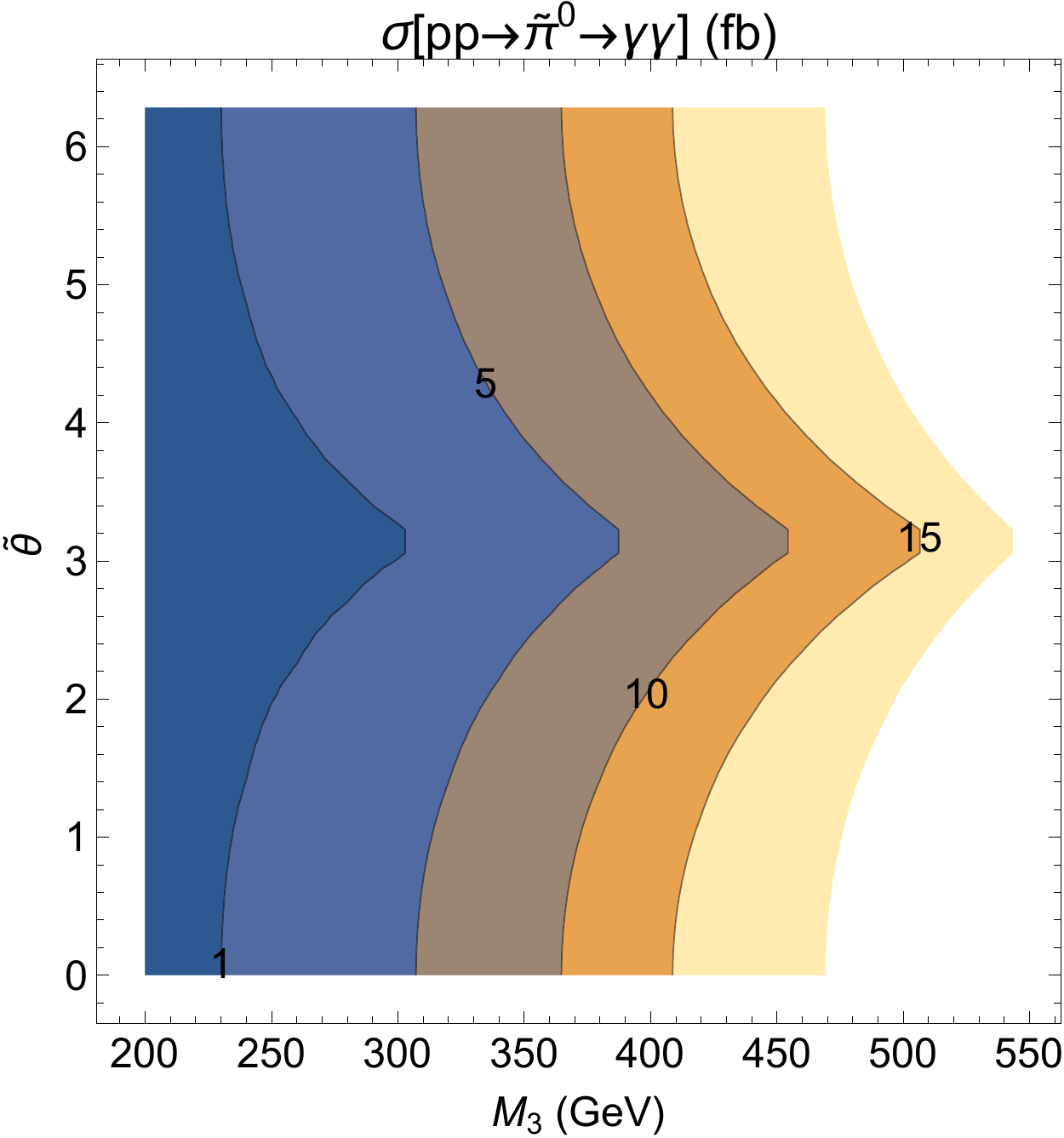}
  \caption{Benchmark model results for the  $pp\rightarrow \hpiN\rightarrow \gamma\gamma$ 13 TeV cross section. In all cases $\fpi$ (and consequently $\tilde\Lambda$) is fixed so that the lightest neutral state has mass $m_{\hpiN}=750$ GeV. We plot on the ($M_3,\thc$) plane, and plots are clipped where any of the masses $M_i$ exceeds half the cutoff $\tilde\Lambda$. First row: $M_1=100$ GeV, $M_2=300$ GeV (left); $M_1=200$ GeV, $M_3-M_2=200$ GeV (right). Second row: $M_1=400$ GeV, $M_3-M_2=100$ GeV (left); $M_2-M_1=M_3-M_2=100$ GeV (right). } 
  \label{fig:xsec750}
\end{figure}

Fig.~\ref{fig:xsec750} shows the $pp\rightarrow \hpiN\rightarrow \gamma\gamma$ cross section at 13 TeV. We compute the cross section at leading order and apply a $K$-factor of 1.6~\cite{Spira:1995rr} in each of the four scenarios. In most cases the rate decreases as $\thc\rightarrow\pi$ due to the increase in $\tilde\Lambda$, which controls the dimension-5 coupling to $\ggt$. In Scenario 2 the rate becomes small and insensitive to $\thc$ at low $M_3$, where $M_2$ is approaching $M_1$ and the mixing angle is suppressed. For fixed $\thc$, the rate in Scenario 1 decreases with increasing $M_3$ again because of mixing angle suppression. In the other scenarios, the rate mostly increases with $M_3$; since two or three masses are becoming large together, $m_{\hpiN}$ increases unless the cutoff is lowered, leading to larger anomaly-type couplings.

We see that on each slice of parameter space, there are sizable regions consistent with the observed diphoton excess of order 1-10 fb, and in particular in each case there are viable regions for all values of $\thc$.


\subsection{Parity violating Hyper-meson Decays}
\label{sec:pvd}
Parity-violating triple-meson couplings can give rise to new decay channels that are absent when $\thc=0$. In the simplest model of one color triplet and one color singlet hyperfermion studied in~\cite{Nakai:2015ptz}, a natural candidate is $\tilde{\eta}^\prime\rightarrow\hpiN\hpiN$ (this channel was also noted recently in~\cite{Harigaya:2016pnu}). This mode is also present in our five-flavor benchmark model when $\thc\neq0$. However, in both cases the coupling can only be studied in chiral perturbation theory at large $N_{\tilde{c}}$, and moreover the $\tilde{\eta}^\prime$ may be too heavy to produce at the LHC if $\tilde{\Lambda}$ is large. 

Alternative channels in our benchmark model are $\heta\rightarrow\hpiN\hpiN$, allowed if $m_{\heta}>2m_{\hpiN}$, and $\heta\rightarrow\hpiN{\tilde{\pi}}^{0\ast}\rightarrow\hpiN gg$, relevant when $m_{\heta}<2m_{\hpiN}$.\footnote{The possibility of these types of decays in the presence of $\thc$ was also noted in~\cite{Franceschini:2015kwy}. We thank Michele Redi for bringing this to our attention.}

Due to the large tree-level cubic coupling $V_{\rm cubic}\supset A\heta\hpiN\hpiN$ (where $A$ is a dimension-1 coefficient, given in Eq.~(\ref{eq:cubics}) in the limit $M_{1,2}\ll M_3$), and the fact that the next 2-body $\heta$ decay mode is the loop-suppressed decay into gluons through~(\ref{eq:etaGGt}), the $\heta\rightarrow\hpiN\hpiN$ channel is expected to be dominant when it is kinematically accessible. The rate for this decay is
\begin{align}
\Gamma_{\heta\to2\hpiN}=\frac{A^2}{32\pi m_{\heta}}\sqrt{1-\frac{4m_{\hpiN}^2}{m_{\heta}^2}}\;.
\end{align}
The subsequent decays of the $\hpiN$ lead to the final states $(gg)(gg)$, $(gg)(\gamma\gamma)$, and $(\gamma\gamma)(\gamma\gamma)$, where the parentheses indicate that the dijets or diphotons reconstruct the $\hpiN$ mass of $750~\rm GeV$. The invariant mass of the two pairs peaks at the $\heta$ mass. Compared to the paired dijets, the $(gg)(\gamma\gamma)$ final state avoids combinatoric backgrounds and offers increased resolution on the $\hpiN$ mass using the diphotons, but has branching fraction suppressed by the electromagnetic coupling. For a detailed discussion of these issues in the context of Higgs boson decays, see~\cite{Curtin:2013fra}.

If  $m_{\heta}<2m_{\hpiN}$, the parity-violating decay is 3-body and is heavily suppressed by the off-shell $\hpiN$. The rate is
\begin{align}
\Gamma_{\heta\to\hpiN gg}=\frac{A^2}{16\pi^2 m_{\hpiN}}\frac{m_{\heta}}{m_{\hpiN}}\frac{\Gamma_{\hpiN\to gg}}{m_{\hpiN}}I\left(\frac{m_{\hpiN}^2}{m_{\heta}^2},\frac{\Gamma_{\hpiN}^2}{m_{\hpiN}^2}\right)
\end{align}
where
\begin{align}
I\left(x,y\right)\equiv\int_0^{\left(1-\sqrt x\right)^2}\!\!\!\! dz\frac{z^2\sqrt{1+x^2+z^2-2x-2z-2xz}}{\left(x-z\right)^2+x^2y}\;.
\end{align}
To get a sense of the magnitude of this suppression, note that as the mass splitting  $\Delta m\equiv m_{\heta}-m_{\hpiN}$ decreases, the 3-body rate falls off rapidly,
\begin{align}
\Gamma_{\heta\to\hpiN gg}\simeq\frac{2A^2}{105\pi^2m_{\heta}}\frac{\Gamma_{\hpiN\to gg}}{m_{\hpiN}}\left(\frac{\Delta m}{m_{\heta}}\right)^7\;,
\end{align}
where we have ignored terms of order $\Gamma_{\hpiN}^2/m_{\hpiN}^2$. 

Because the 3-body decay is generally negligible, it is most interesting to focus on cases where the 2-body decays to on-shell $\hpiN$'s can proceed in regions of parameter space overlapping with a $\hpiN\rightarrow\gamma\gamma$ cross section compatible with the observed excess. Comparison of Figs.~\ref{fig:meta} and~\ref{fig:xsec750} indicates this overlap is most likely to occur in scenarios 2 and 3, where the relationship $M_2\simeq M_3$ leads to large mixing between the $\Pi$ states. 

It is worth a note of explanation why we have not taken the strict ``simplifying" limit $M_2=M_3$ in scenarios 2 and 3. In the exact $M_2=M_3$ limit, for $M_1<M_3$, the $\heta$ carries charge -1 under a discrete symmetry which is a hypercolor analog of $\cal G$-parity in QCD (for further discussion and application of such symmetries, see~\cite{Bai:2010qg,Bai:2015nbs}). Thus this ``isospin-like" limit forbids couplings in the hyperpion potential with an odd number of $\heta$ particles, including the cubic coupling $\heta\hpiN\hpiN$ we wish to study. Therefore, we keep a modest $M_3-M_2$ splitting in the benchmarks.

\begin{figure}[t]
  \centering
 \includegraphics[height=0.5\textwidth]{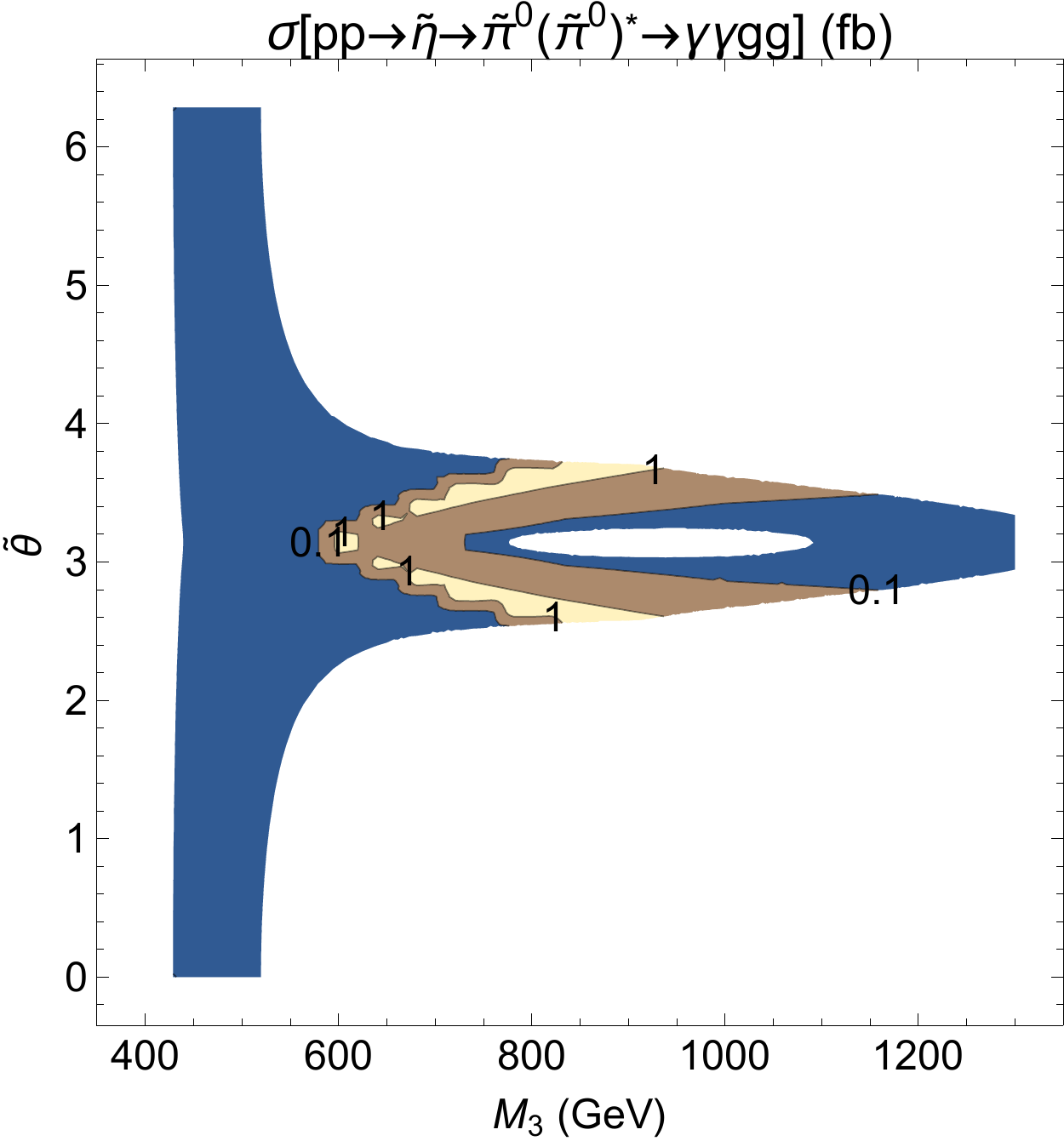}\;\;\;\;\;\;
  \includegraphics[height=0.5\textwidth]{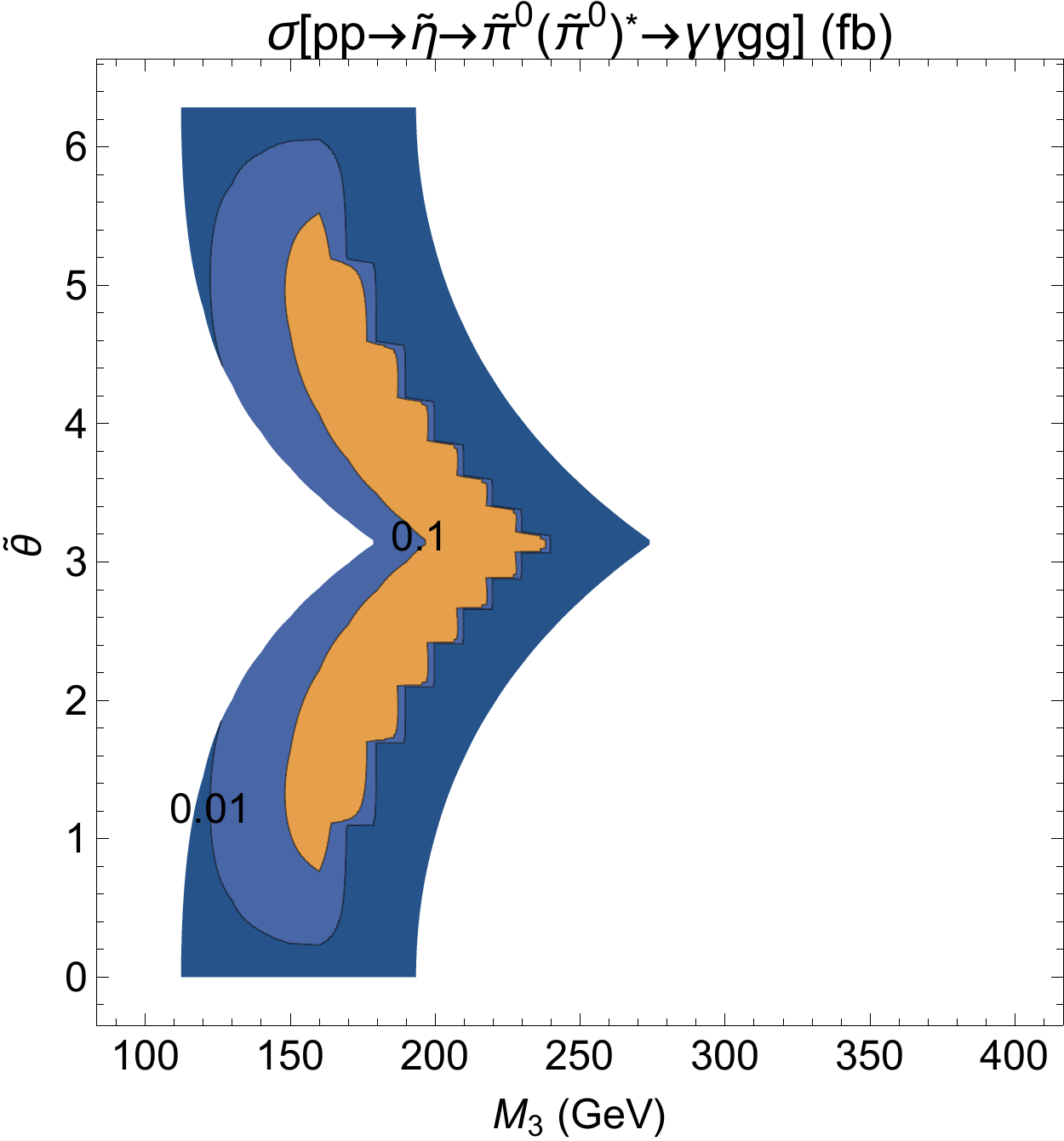}\\
  \caption{Benchmark model results for the $pp\rightarrow \heta\rightarrow \hpiN\hpiN\rightarrow \gamma\gamma gg$ 13 TeV cross section. Panels correspond to Scenarios 2 ($M_1=200$ GeV, $M_3-M_2=200$ GeV; left) and 3 ($M_1=400$ GeV, $M_3-M_2=100$ GeV; right). We have only shaded regions of parameter space where $\sigma(pp\rightarrow\hpiN\rightarrow\gamma\gamma)$ falls in the range 1-10 fb.}
  \label{fig:hetaprod}
\end{figure}

In Fig.~\ref{fig:hetaprod} we plot the 13 TeV cross sections for $pp\rightarrow\heta\rightarrow\gamma\gamma gg$ in these scenarios,  restricting the plots to points where $\sigma(pp\rightarrow\hpiN\rightarrow\gamma\gamma)$ is in the range 1-10 fb. We find that in the parameter space consistent with the diphoton excess, there are sizable regions in which $pp\rightarrow\heta\rightarrow\gamma\gamma gg$ may be observable at the LHC with $\cal O$(100) fb$^{-1}$ of integrated luminosity.

\subsection{Other $\thc$-dependent Phenomenology}

We conclude our discussion of the benchmark model with some brief comments on other probes of $\thc$.

Apart from parity-violating hyperpion decays, another test of nonzero $\thc$ at colliders arises in the angular distribution of leptons in the decay mode $\hpiN\rightarrow ZZ$ when both $Z$'s decay leptonically, as has been discussed in the case of the Higgs (see, e.g.,~\cite{Chen:2014gka}). In principle, such a measurement could determine the parity-violating coupling of the $\hpiN$ to a pair of $Z$'s. In the benchmark model we have described, this coupling comes from the term $\thc\,\hpiN B^{\mu\nu}B_{\mu\nu}$ where $B^{\mu\nu}$ is the $U(1)_Y$ field strength, in analogy with Eq.~(\ref{eq:piGG}). As mentioned above, this parity-violating $\hpiN$ coupling to the gauge boson kinetic term is chirally suppressed relative to the parity-conserving coupling to the topological charge density (corresponding to the fact that $\thc$ is unphysical in the limit of a vanishing hyperquark mass).\footnote{The parity-violating triple pion couplings are also chirally suppressed, but in contrast, the decays to which they give rise are a leading-order effect.} Therefore, sensitivity must be high to disentangle the subleading contribution. Furthermore, this measurement is challenging because of the small branching ratio for leptonic $Z$ decays.

Parity violation also permits the hyperpion states to mix with the Higgs boson, which in principle might be observable at colliders through new $\hpiN$ decay channels. This mixing can be generated in the UV by dimension-5 $\bar\psi\psi |H|^2$ operators, or in the IR through gauge boson loops sensitive to $\thc$. The former contributions are model-dependent and may be negligible if the scale suppressing the higher-dimension operators is large, while the latter appear at 3-loop order in our benchmark model and are likewise negligible. Therefore, at least in models of the type studied here, Higgs mixing is expected to be unobservable in practice.

A potential low-energy probe of parity violation comes from searches for the neutron electric dipole moment (EDM). As pointed out in the appendix of~\cite{Harigaya:2016pnu}, the leading contribution comes from the generation of the three-gluon Weinberg operator~\cite{Weinberg:1989dx} and could plausibly be tested by next generation searches. Direct contributions to quark (chromo)EDMs from diagrams involving $\hpiN$ exchange occur at two loops, but are effectively four-loop in magnitude since the couplings to gluons and photons are generated at one loop, and are thus well below current or near future experimental sensitivity.

\section{Strong CP}
\label{strongcp}
In VC models, the hypercolor sector typically gives new ${\cal O}(1)$ contributions to $\tqcd$. In this section we illustrate the shift in $\tqcd$ in the benchmark model and discuss the implications of this effect for solutions to the strong CP problem.

\subsection{$\thc$ and $\tqcd$}

We have already seen an obvious contribution to $\tqcd$, Eq.~(\ref{eq:dtqcd1}), from the new set of quarks $\psi_3$. There is also a contribution from the $\hpiB$ vev. 
From its coupling to the QCD topological charge density, Eq.~(\ref{eq:etaGGt}), we see that $\hpiB$ gives a threshold correction to $\tqcd$,
\begin{align}
(\Delta\tqcd)_2=\frac{N_{\tilde{c}}}{\sqrt{15}}\frac{\langle \hpiB\rangle}{\fpi}\;.
\label{eq:dtqcd2}
\end{align}
Together, the two contributions give a total shift in $\tqcd$ of 
\begin{align}
\Delta\tqcd=N_{\tilde{c}}\left[\frac{\phi_3}{3}-\frac{M_1M_2\sin(\thc)}{2M_3\sqrt{M_1^2+M_2^2+2M_1M_2\cos(\thc)}}\right]\;.
\label{eq:deltatheta}
\end{align}
in the limit studied in Sec.~\ref{sec:analyt} and using Eq.~(\ref{eq:veta}). More generally, we expect $\Delta\tqcd$ to receive the shift from $\phi_3$, as well as a dynamical shift of order $\thc$ that reduces to the second term in Eq.~(\ref{eq:deltatheta}) in the appropriate limit.

In the absence of other sources of chirality violation, the phase $\phi_3$ can always be moved completely into the QCD $\tqcd$ angle. However even if $\phi_3=0$, there is still an independent contribution to $\tqcd$ from $\thc$, and as we have discussed, $\thc$ has in principle a number of other observable effects.
The $\thc$ contribution is generic, although it arises in different ways in different models. For example, in the minimal model of~\cite{Nakai:2015ptz}, there is no $\heta$ state, but the $\hpiN$ couples directly to the QCD anomaly and carries a vev in the presence of $\thc$. In both cases, the effect is unsuppressed by couplings or loop factors. 

Crudely speaking, a shift in $\tqcd$ near the TeV scale tells us that the strong CP problem ``has yet to be solved" by dynamics at lower energies. We will make this assertion more precise below.

\subsection{Solutions to Strong CP: UV vs. IR}
Proposed solutions to the strong CP problem fall broadly into two categories. The first type of solution deals with infrared physics, and leaves infrared signatures of its presence. Two examples are the Peccei-Quinn (PQ) solution and its signature axion, and the $m_u=0$ solution, disfavored by lattice data. The second type of solution uses ultraviolet physics (compared to QCD), and from the infrared point of view, largely appears to be a miracle. Examples include the Nelson-Barr (NB) models of spontaneous CP violation~\cite{nelsoncp,barrcp,barrcp2,Bento:1991ez}, left-right models with spontaneous P violation~\cite{Beg:1978mt,mohapatrasenjanovic,Georgi:1978xz,Babu:1989rb,barrsenjanovic,Mohapatra:1995xd, Kuchimanchi:1995rp,Mohapatra:1996vg,Kuchimanchi:2010xs,Mohapatra:1997su}, and models with a new massless colored fermion confined by a new gauge group~\cite{hook_cp_violation}.

Although axions are very weakly coupled, there exist a variety of experimental probes. 
Comparatively, UV solutions to strong CP are usually difficult to test. In general the relevant scales can be vastly higher than the TeV scale, leaving little trace at low energies, other than the value of $\tqcd$. In specific cases, some new states might be accessible at colliders~\cite{barrsenjanovic,hook_cp_violation}.   Alternatively, the study of phases in other new TeV-scale dynamics, such as $\thc$ in VC models, might be used to discriminate whether strong CP is solved by ultraviolet or infrared physics.

Solving strong CP in the UV is a delicate matter: it relies on the curious fact that the renormalization of $\tqcd$ within the SM alone is tiny. If microscopic physics with scale $\Lambda_{UV}$ can explain why $\tqcd=0$ is the right UV boundary condition for the EFT below $\Lambda_{UV}$, then as long as the EFT is not too different from the SM, $\tqcd\approx 0$ will be preserved. On the other hand, if there is still substantial BSM physics below $\Lambda_{UV}$, it can easily spoil the solution to strong CP through radiative contributions to $\tqcd$.\footnote{Threshold corrections to $\tqcd$ at $\Lambda_{UV}$, including from whatever dynamics stabilizes $\Lambda_{UV}/M_p$, present additional theoretical constraints on UV solutions to strong CP~\cite{dinedraperbn,Albaid:2015axa}.}
The detection of new pion-like states coupling to gluons, and a large new vacuum angle $\thc$, is a clear example: threshold corrections like Eq.~(\ref{eq:deltatheta}) generically provide a large shift in $\tqcd$. The threshold correction is innocuous if $\thc$ itself is tiny, but then we must solve a {\it second} strong CP problem. From a model-building perspective, this is most natural if $\tqcd$ and $\thc$ are suppressed in the UV by the same mechanism.

Below we briefly review specific UV and IR solutions to the strong CP problem and how they are affected by the addition of a VC sector. 

~ \\
\noindent {\bf Nelson-Barr} \\
In NB models~\cite{nelsoncp,barrcp,barrcp2,Bento:1991ez}, CP is taken to be a good underlying symmetry, so in the ultraviolet $\tqcd$ and $\thc$ both vanish by assumption. Since CP must be broken at low energies, a sector is added to spontaneously break it at some intermediate scale $M_{CP}$ through a complex vev for a field $\sigma$ (in general, a set of fields).  The particle content, interactions, and symmetries are arranged so that the $\sigma$ vev is communicated to the CKM phase in an unsuppressed way, while $\tqcd$ is not generated, at least at tree level. 

Without specifying the full structure of the NB sector, let us add a VC sector near the TeV scale and take $M_{CP}>{\rm TeV}$. In the absence of additional symmetries on the VC sector, $\thc$ is generated when CP is spontaneously broken, for example, by renormalizable couplings of the form 
\begin{align}
{\cal L}\supset f_{ij} \sigma \bar\psi_i\psi_j+h.c.\nonumber\\
\Rightarrow \Delta\thc\sim{\rm arg}(\sigma)
\label{eq:sigmapsipsiNB}
\end{align}
In this case, $\thc$ feeds in to $\tqcd$ in an ${\cal O}(1)$ way near the TeV scale, reintroducing the strong CP problem. 

On the other hand, it is not difficult to forbid couplings like~(\ref{eq:sigmapsipsiNB}) with discrete symmetries, for example a $\mathbb{Z}_2$ under which $\sigma$, $\psi$, and $\bar\psi_i$ are all odd. Indeed, such symmetries are a necessary ingredient of NB models, even without VC sectors, in order to forbid other problematic renormalizable couplings involving $\sigma$. The symmetries may be extended to the hyperfermion couplings, and it is conceivable that $\thc$ is sufficiently small at $M_{CP}$ to preserve the NB solution. 

We will not attempt to build a complete model exhibiting both NB and VC sectors here, but simply note that the presence of the VC sector in NB models requires $\thc$ to be as well-protected as $\tqcd$. A signature of this case is that $\thc$ will not be observable. If, on the other hand, $\thc$ is observed, we may conclude that strong CP must solved in another, more infrared way.

~ \\
\noindent {\bf Parity Models} \\
A similar but distinct class of UV solutions to strong CP, based on parity, was first studied in~\cite{Beg:1978mt,mohapatrasenjanovic,Georgi:1978xz,Babu:1989rb,barrsenjanovic}. In these models, a (generalized) parity symmetry is enforced in the ultraviolet theory. The simplest implementation expands the SM gauge group to $SU(3)\times SU(2)_L\times SU(2)_R\times U(1)_Y$, and parity exchanges $SU(2)_L\leftrightarrow SU(2)_R$~\cite{barrsenjanovic} (see also the recent study~\cite{hook_p_violation}). To symmetrize the fermion content, mirror fermions are added. For example, the ordinary left-handed electroweak doublet $Q$, transforming as $(3,2,1,1/6)$, is matched with a mirror left-handed field $\bar Q^\prime$ transforming as $(\bar 3, 1, 2, -1/6)$. Parity then exchanges $Q\leftrightarrow {\bar Q}^{\prime*}$, and requires $\tqcd=0$ in the UV. In the VC extensions of the SM like the benchmark model studied here, parity can act on the new vectorlike hyperfermions as $\psi_i\leftrightarrow \bar \psi_i^*$. With this transformation, parity also requires $\thc=0$ in the UV. 

Like CP in NB models, parity must be spontaneously broken at low scales. Again there are typically couplings that reintroduce $\thc$ at tree level. For example, if parity is broken by a vev for a pseudoscalar $a$, then
\begin{align}
{\cal L}\supset i y_{ij} a (\bar\psi_i\psi_j-\bar\psi_j^*\psi_i^*)
\label{eq:apsipsi}
\end{align}
is parity-invariant for hermitian $y$ and contributes to $\thc$ when parity is broken. 

We draw the same conclusion as in the case of NB: the presence of the VC sector in left-right models requires $\thc$ to be as well-protected as $\tqcd$, which is plausible, at least at tree-level, with the addition of symmetries to forbid couplings between the hyperfermions and the parity-breaking sector. $\thc$ will not be observable if such models are realized in nature.\footnote{For recent studies of left-right solutions to strong CP in the context of the diphoton excess, see~\cite{Cao:2015xjz,Dev:2015vjd}.} If, on the other hand, $\thc$ is observed, we conclude as before that strong CP is not solved by a P symmetry of the UV theory.

~ \\
\noindent {\bf The QCD Axion} \\
We conclude this section by commenting briefly on the most plausible IR solution to strong CP, the Peccei-Quinn mechanism~\cite{Peccei:1977hh,Peccei:1977ur,Weinberg:1977ma,Wilczek:1977pj}, and its interplay with a new hypercolor sector. 

As we have emphasized, unlike the UV solutions to strong CP, the cancellation of $\tqcd$ by a vev for an axion coupling to $\ggt$ is unspoiled by any threshold corrections to $\tqcd$ down to very low scales. The vacua of the QCD-induced axion potential are simply shifted to relax whatever value $\tqcd$ takes in the IR.

This is not to say, however, that the Peccei-Quinn solution is automatic in any theory with an axion, a hypercolor sector, and a generic value of $\thc$. 
If PQ symmetry is anomalous under hypercolor, there is a new contribution to the axion potential that swamps the QCD contribution,
\begin{align}
\frac{M\tilde\Lambda^3}{m \Lambda^3}\sim 10^{16}\;,
\end{align}
resulting in the relaxation of $\thc$, but not $\tqcd$. In the case of field theory axions, avoiding this contribution amounts to constraints on the field content such that the PQ anomaly with hypercolor vanishes.

\subsection{New Massless Quarks}
\begin{figure}[t]
  \centering
 \includegraphics[height=0.45\textwidth]{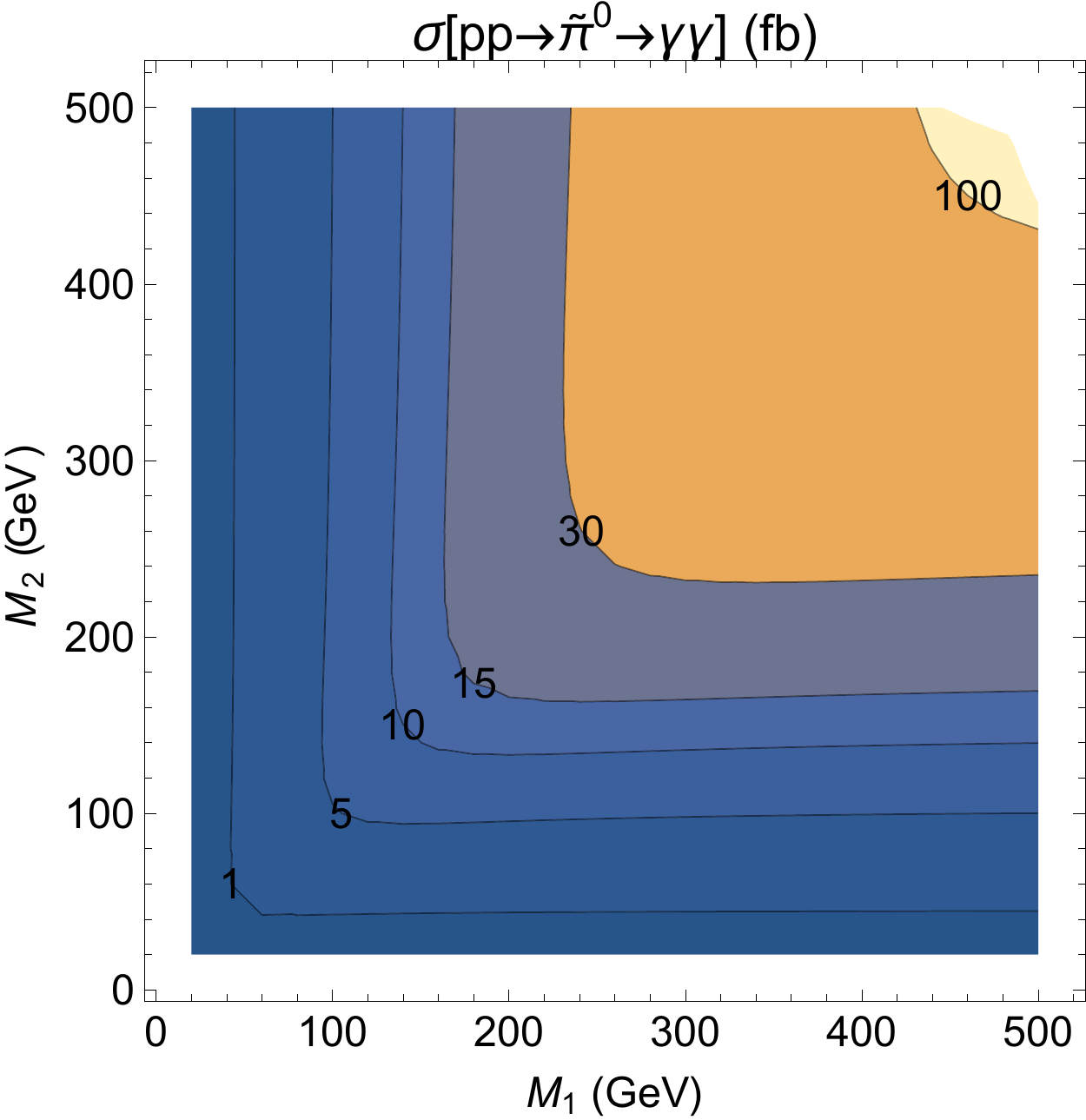}
  \caption{Benchmark model results on the $(M_1,M_2)$ plane for the $pp\rightarrow  \hpiN\hpiN\rightarrow \gamma\gamma $ 13 TeV cross section in the case $M_3=0$. In contrast to previous parameter regimes, here $\thc$ is unphysical.}
  \label{fig:M30}
\end{figure}
There is another category of solutions to the strong CP problem that bears mention and does not fall neatly into the UV/IR classification discussed above: the possibility that the sector responsible for the diphoton resonance is itself complicit in the solution to strong CP. In the case of a new strongly coupled hypercolor sector, an example of such a solution can arise when there is a new massless colored and hypercolored quark.

A model of this type was studied prior to the diphoton excess in Ref.~\cite{hook_cp_violation}, which proposed that strong CP might be solved if there is a new $\tilde N=3$ hypercolor sector, a color- and hypercolor-fundamental fermion with vanishing mass, and a $\mathbb{Z}_2$ mirror symmetry that fixes $\tqcd=\thc$ to high precision. Then, the same anomalous chiral rotation may be used to simultaneously eliminate $\thc$ and $\tqcd$ from the theory. The $\mathbb{Z}_2$ is spontaneously broken at very high scales by a very large vev for the mirror Higgs field, so that the mirror partners of the SM fermions are all very heavy and the hypercolor group runs strong before QCD. From a top-down perspective such models appear to face fine-tuning challenges~\cite{Albaid:2015axa}, but from a bottom-up point of view it is interesting to study their compatibility with the diphoton excess and their further predictions. 

At low energies, the field content of our benchmark model and the parameter limits $\thc\rightarrow\tqcd$, $M_3\rightarrow 0$ are almost sufficient to realize the structure required of this type of solution. In addition, we have to add the mirror partners of the color singlet hyperfermions, which become ordinary hypercolor-singlet vectorlike quarks in the fundamental of QCD, with masses set by $M_1$ and $M_2$. These degrees are freedom are likely to be long-lived, since higher-dimension operators must be added to permit their decay.

In Fig.~\ref{fig:M30} we plot the diphoton cross section in the $M_3\rightarrow 0$ limit. We see that compatibility with the observed rate indicates an ${\cal O}(100)$ GeV mass for one of the two singlet hyperfermions. Consequently, the model predicts a new light colored vectorlike fermion in addition to the hyperpion sector. The phenomenology of this state is model-dependent due to the freedom in the extra structure that must be added to allow it to decay, but it is likely to be severely constrained.


\section{Conclusions}
\label{concl}

New QCD-like sectors provide attractive and natural candidates for the diphoton excess observed by ATLAS and CMS. In the presence of light fermions charged under the new strong gauge group, a neutral composite pseudo-Goldstone state $\hpiN$, analogous to the $\pi^0$, may couple to QCD and QED through chiral anomalies.  However, unlike ordinary QCD, the new sector may exhibit strong parity violation through a large vacuum angle $\thc$. We have studied the impact of $\thc$ on the physics of the new pseudo-Goldstone sector and the importance of $\thc$ as a probe of the strong CP problem. 

Varying $\thc$ reveals a rich vacuum structure and has substantial impact on the pseudo-Goldstone spectrum. Furthermore, in all models of this type, $\thc$ controls parity violating decays of the form $\hetap\rightarrow\hpiN\hpiN$. However, because of the axial anomaly in the new sector, these decays are typically not calculable in chiral perturbation theory. We have instead considered a larger benchmark model with an additional pseudo-Goldstone state $\heta$, analogous to the $\eta$ of QCD, and studied the process $\heta\rightarrow\hpiN\hpiN$ in ChPT.  We find that in the benchmark model, the $gg\gamma\gamma$ final state for this process can be probed at the LHC in sizable regions of parameter space consistent with the diphoton excess.

There are other potential experimental probes of $\thc$ deserving of dedicated analysis, in particular whether angular distributions in $\hpiN\rightarrow ZZ\rightarrow 4\ell$ offer sufficient sensitivity to disentangle the CP-conserving from the CP-violating contributions. We reserve this question for future work.

Chiral anomalies with QCD allow resonant production of the new pseudo-Goldstones at the LHC through gluon fusion, $pp\rightarrow \hpiN,\heta$. The same couplings generate threshold corrections to $\tqcd$ of order $\thc$ near the TeV scale. Thus, $\thc$ is an efficient discriminator of whether the strong CP problem is solved by ultraviolet or infrared physics. If $\thc$ is small, the most plausible explanation is that a UV symmetry like P or CP protects both $\tqcd$ and $\thc$. If, on the other hand, $\thc$ is large, the threshold correction implies that $\tqcd$ must be eliminated by an IR mechanism like the axion.

\section*{Acknowledgements}
PD thanks the Santa Cruz Institute for Particle Physics for hospitality during the course of this work. The work of DM was supported in part by the U.S. Department of Energy under Grant No. DE-SC0011637.

\bibliography{parity_in_VC_refs}
\bibliographystyle{jhep}

\end{document}